\documentclass[11pt,a4paper]{article}
\pdfoutput=1
\usepackage{jheppub}
\usepackage{ifpdf}
\usepackage{afterpage}
\usepackage{amssymb}
\usepackage{amsmath}
\usepackage{graphicx}
\usepackage{longtable}
\usepackage{verbatim}
\usepackage{amsfonts}
\usepackage{bbold}
\usepackage[table,usenames,dvipsnames]{xcolor}
\usepackage{titlesec}
\usepackage{cancel}
\usepackage{enumitem}
\usepackage{hyperref}
\usepackage{soul}

\usepackage{tikz}
\usetikzlibrary{trees}
\usetikzlibrary{decorations.pathmorphing}
\usetikzlibrary{decorations.markings}
\tikzset{
    photon/.style={decorate, decoration={snake}, draw=black},
    wino/.style={draw=redwine},
    electron/.style={draw=black, postaction={decorate},
        decoration={markings,mark=at position .55 with {\arrow[draw=black]{>}}}},
    scalar/.style={draw=black, dashed,postaction={decorate},
        decoration={markings,mark=at position .55 with {\arrow[draw=black]{>}}}},
    gluon/.style={decorate, draw=black,
        decoration={coil,amplitude=4pt, segment length=5pt}}
}

\newcommand{\slashed}[1]{\displaystyle{\not}#1}

\newcommand{\bear}{\begin{array}}
\newcommand{\ear}{\end{array}}
\newcommand{\beq}{\begin{equation}}
\newcommand{\eeq}{\end{equation}}
\newcommand{\beqa}{\begin{eqnarray}}
\newcommand{\eeqa}{\end{eqnarray}}

\def\OMIT#1{{}}
\newcommand{\lsim}{\mathrel{\rlap{\lower4pt\hbox{\hskip1pt$\sim$}}
    \raise1pt\hbox{$<$}}}         
\newcommand{\gsim}{\mathrel{\rlap{\lower4pt\hbox{\hskip1pt$\sim$}}
    \raise1pt\hbox{$>$}}}         

\newcommand{\ignore}[1]{}

\begin{document}
\title{\boldmath New inflationary probes of axion dark matter}
\author[a]{Xingang Chen,}
\author[b,c]{JiJi Fan,}
\author[b]{Lingfeng Li}
\affiliation[a]{Institute for Theory and Computation, Harvard-Smithsonian Center for Astrophysics, 60 Garden Street, Cambrige, MA, 02138, USA}
\affiliation[b]{Department of Physics, Brown University, Providence, RI, 02912, USA}
\affiliation[c]{Brown Theoretical Physics Center, Brown University, Providence, RI, 02912, USA}
\emailAdd{xingang.chen@cfa.harvard.edu}
\emailAdd{jiji\_fan@brown.edu}
\emailAdd{lingfeng\_li@brown.edu}
\vspace*{1cm}

\abstract{If a light axion is present during inflation and becomes part of dark matter afterwards, its quantum fluctuations contribute to dark matter isocurvature. In this article, we introduce a whole new suite of cosmological observables for axion isocurvature, which could help test the presence of axions, as well as its coupling to the inflaton and other heavy spectator fields during inflation such as the radial mode of the Peccei-Quinn field. They include correlated clock signals in the curvature and isocurvature spectra, and mixed cosmological-collider non-Gaussianities involving both curvature and isocurvature fluctuations with shapes and running unconstrained by the current data analyses. Taking into account of the existing strong constraints on axion isocurvature fluctuations from the CMB, these novel signals could still be sizable and potentially observable. In some models, the signals, if observed, could even help us significantly narrow down the range of the inflationary Hubble scale, a crucial parameter difficult to be determined in general, independent of the tensor mode.  }
\maketitle
\flushbottom

\section{Introduction}
\label{sec:intro}

An axion is a pseudo-scalar field that has a discrete shift symmetry, $a \cong a + 2\pi f_a$, with $f_a$ known as the decay constant. Despite its apparent simplicity,  an axion could have a plethora of phenomenological and cosmological applications, ranging from solving the strong CP problem in the Standard Model (SM) of particle physics~\cite{Peccei:1977hh, Peccei:1977ur,Weinberg:1977ma,Wilczek:1977pj}, to serving as a classic cold dark matter (DM) candidate~\cite{Preskill:1982cy,Dine:1982ah,Abbott:1982af}. It could be realized naturally as a light pseudo-Nambu-Goldstone boson from the spontaneous breaking of a global Abelian symmetry, the Peccei-Quinn (PQ) symmetry, $U(1)_{\rm PQ}$, at the high energy scale $f_a$. Though the axion scenario was proposed more than 40 years ago~\cite{Peccei:1977hh, Peccei:1977ur,Weinberg:1977ma,Wilczek:1977pj}, it continues to serve as one of the most motivated feebly-coupled particles beyond the SM, generating growing theoretical and experimental studies. For some recent reviews on axions as well as discussions of future plans and open questions, see~\cite{Sikivie:2006ni,Marsh:2015xka, DiLuzio:2020wdo, Adams:2022pbo, Agrawal:2022yvu, Asadi:2022njl}. 

A light axion, either a QCD axion or a general axion-like particle, could have a close interplay with the inflationary paradigm, one leading explanation for the origin of our observable universe~\cite{Guth:1980zm, Starobinsky:1980te, Linde:1981mu,Albrecht:1982wi}. In particular, if $U(1)_{\rm PQ}$ is spontaneously broken during inflation, the resulting massless axion would be present and undergoes random quantum fluctuations with an amplitude set by the inflationary Hubble scale, $H$~\cite{Steinhardt:1983ia, Seckel:1985tj, Lyth:1989pb, Turner:1990uz, Linde:1991km}. These fluctuations, usually referred to as axion isocurvature perturbations, are independent of the primordial inflaton fluctuations. After inflation, they could be imprinted on the DM perturbations, once the axion becomes (part of) DM, while the inflaton perturbations are inherited by radiation, baryons and other DM components. Given the current non-observation of isocurvature modes in the cosmic microwave background (CMB) observations, there exists a strong constraint on the combination of the axion DM fraction and the size of the axion isocurvature perturbations~\cite{Planck:2018jri}. In the QCD axion scenario, the constraint is often interpreted as the ``axion isocurvature problem": inflationary QCD axion DM, if it is the leading component of DM, is incompatible with high-scale inflation (e.g., $f_a \sim 10^{11}$ GeV, which will make the relic abundance of QCD axion DM through the minimal misalignment mechanism match the observed DM abundance, requires the inflationary Hubble scale to be below $10^7$ GeV~\cite{Planck:2018jri}). In this paper, we will not intend to provide a new solution to this problem, remedying the relationship between the QCD axion DM and high-scale inflation. Instead, we take a different perspective and address the following question: given the existing constraint on the size of axion isocurvature perturbation, could there exist other non-trivial cosmological observables testing the existence of light axions during inflation?   

To answer this question, we make use of the recent developments in cosmological collider physics and primordial standard clock models, which provide new opportunities for probing high energy physics during inflation. 

During inflation, particles with masses around and below the Hubble scale are produced on-shell. It has been pointed out that these heavy spectator particles could generate non-analytical momentum dependence and angular dependence of the primordial non-Gaussianities (three- and higher-point correlators) ~\cite{Chen:2009zp,Arkani-Hamed:2015bza}, due to their quantum motions. Observations of primordial non-Gaussiantity (NG) could provide rich information about the mass and spin of the heavy particles. This program, by its analogy to the mass-spin measurements at terrestrial particle colliders, has been dubbed ``cosmological collider" (CC) physics. It has attracted rapidly growing attention~\cite{Chen:2009we,Chen:2009zp,Baumann:2011nk,Assassi:2012zq,Chen:2012ge,Sefusatti:2012ye,Norena:2012yi, Pi:2012gf,Noumi:2012vr,Gong:2013sma,Emami:2013lma, Arkani-Hamed:2015bza,Chen:2015lza, Dimastrogiovanni:2015pla,Kehagias:2015jha,Chen:2016nrs,Lee:2016vti,Meerburg:2016zdz,Chen:2016uwp,Chen:2016hrz,Chen:2017ryl,Kehagias:2017cym,An:2017hlx,An:2017rwo,Iyer:2017qzw,Kumar:2017ecc,Chen:2018sce,Chen:2018xck,Chua:2018dqh,Kumar:2018jxz,Wu:2018lmx,MoradinezhadDizgah:2018ssw,Saito:2018omt,Tong:2018tqf,Fan:2019udt,Alexander:2019vtb,Lu:2019tjj,Hook:2019zxa,Hook:2019vcn,Kumar:2019ebj,Wang:2019gbi,Liu:2019fag,Wang:2019gok,Wang:2020uic,Li:2020xwr,Wang:2020ioa,Fan:2020xgh,Kogai:2020vzz,Bodas:2020yho,Aoki:2020zbj,Arkani-Hamed:2018kmz,Baumann:2019oyu,Baumann:2020dch,Maru:2021ezc,Lu:2021gso, Lu:2021wxu,Wang:2021qez,Tong:2021wai,Pinol:2021aun, Cui:2021iie,Tong:2022cdz,Reece:2022soh,Pimentel:2022fsc,Qin:2022lva,Chen:2022vzh,Jazayeri:2022kjy, Ghosh:2022cny, Qin:2022fbv, Xianyu:2022jwk,Niu:2022quw, Niu:2022fki,Wang:2022eop,Aoki:2023tjm,Werth:2023pfl}, given its tremendous potential to study new physics at a much higher energy scale, which could never be produced at a particle collider. 

When sharp features are present in the inflaton trajectory (see \cite{Peiris:2003ff,Adams:2001vc,Bean:2008na, Mortonson:2009qv, Hazra:2010ve, Hazra:2014goa, Miranda:2014fwa, Braglia:2021sun, Braglia:2021rej, Braglia:2022ftm} for motivations from CMB data analyses and see~\cite{Chen:2010xka, Chluba:2015bqa, Slosar:2019gvt} for reviews on primordial features), the energy scale of the cosmological collider can become even orders-of-magnitude higher.
In the primordial standard clock models~\cite{Chen:2011zf, Chen:2011tu, Chen:2012ja, Battefeld:2013xka, Gao:2013ota, Noumi:2013cfa, Saito:2012pd, Saito:2013aqa, Chen:2014joa, Chen:2014cwa, Huang:2016quc, Domenech:2018bnf, Braglia:2021ckn, Braglia:2021sun, Braglia:2021rej, Bodas:2022zca}, the inflaton deviates from the attractor solution temporarily due to some sharp feature in the trajectory. The departure could trigger the oscillation of a massive spectator field with a mass $m \gg H$. As a consequence, in the primordial spectra (two- and higher-point correlators) of the density perturbations, this oscillation gives rise to a clock signal with the frequency sensitive to the background scale factor evolution $a(t)$ and the particle mass $m$ in Hubble unit, providing a way to probe these two kinds of information. The energy injected by the features can also excite quantum oscillations of heavy fields with masses $m \gg H$, again leaving their signatures in the primordial non-Gaussianities \cite{Chen:2022vzh}. 

In this article, we show that due to interactions between the inflaton and the PQ scalar field, whose phase becomes the axion after the breaking of $U(1)_{\rm PQ}$, there arises a whole suite of novel cosmological observables, that allow us to test the existence of the axion during inflation as well as its properties such as the associated PQ breaking scale during inflation. They include correlated clock signals in the curvature and isocurvature spectra; and large (correlated) CC NG signatures in the bispectra involving isocurvature modes with shapes and running that are unconstrained in data analyses so far. These bispectra include the mixed ones between curvature mode(s) and one or two isocurvature modes and pure isocurvature bispectra, with all three modes being the isocurvature ones. As far as we know, Ref.~\cite{Lu:2021gso} is the only paper that studied the CC signal in an inflationary axion model in the pure isocurvature channel. Early papers on axion isocurvature NG but not involving the production of heavy particles with $m \gtrsim H$ as in the CC physics include~\cite{Kawasaki:2008sn, Hikage:2012be}. In \cite{Lu:2021gso}, to achieve a sizable CC signal in that model, the axion needs to have a non-zero classical rolling speed, or in other words, the axion needs to have a mass about the inflationary Hubble scale instead of being effectively massless during inflation. While this is plausible through model building, we will not pursue it in our paper. Instead, we explore a wider range of light inflationary axion models with various couplings and show that with either classical primordial features or chemical-potential type coupling, potentially observable new classical or quantum signals in either power spectra or bispectra could arise, in the parameter space consistent with current observations. 

The paper is organized as follows. In Sec.~\ref{sec:review}, we will review the basics of axion DM isocurvature. In Sec.~\ref{sec:operators}, we will provide a summary of the three models based on effective operators and their main signatures. In the following three sections, Sec.~\ref{sec:model1}, \ref{sec:model2} and \ref{sec:model3}, we present details of each model and relevant key computations. We conclude and point out some future directions in Sec.~\ref{sec:outlook}.

\section{Review of axion dark matter isocurvature}
\label{sec:review}
During inflation, if PQ symmetry is spontaneously broken, a massless axion, denoted by $a$ or equivalently $\theta=a/f_I$ with $f_I$ the PQ field's vacuum expectation value (VEV) during inflation, is generated. The condition for symmetry breaking requires $f_I \gg H$, the Hubble scale during inflation. The amplitude of its fluctuation
\begin{equation}
   \delta \theta = \theta - \langle \theta \rangle = \theta-\theta_i~, 
\end{equation}
with $\langle\theta \rangle = \theta_i$ being the initial misalignment angle, is set by the Gibbons-Hawing temperature~\cite{Gibbons:1977mu}:  
\begin{equation}
~\sigma_\theta = \sqrt{(\delta \theta)^2} = \frac{H}{2\pi f_I}~.
\end{equation}
This fluctuation could be independent of that of the inflaton and leads to the DM isocurvature perturbation once the axion becomes (part of) cold DM after inflation. The isocurvature fluctuation in DM $S_d$ and the contribution due to the axion component $S_a$ are defined by
\begin{equation}
S_d \equiv \frac{\delta \Omega_d}{\Omega_d}-\frac{3}{4}\frac{\delta \Omega_{\rm rad}}{\Omega_{\rm rad}}~,~S_a \equiv \frac{\delta \Omega_a}{\Omega_a}-\frac{3}{4}\frac{\delta \Omega_{\rm rad}}{\Omega_{\rm rad}}~,
\end{equation} 
where $\delta$ denotes fluctuation over the average value, while the subscript rad means photon radiation. The isocurvature fluctuation in DM due to the axion component is given by~\cite{Hertzberg:2008wr,Dine:2017swf} 
\begin{eqnarray}
\label{eq:isocurvatureanda}
     S_d &= & \gamma S_a = \gamma \frac{\theta^2-\langle \theta^2 \rangle}{\langle \theta^2 \rangle}\approx  \frac{2 \gamma  \delta \theta}{\theta_i} =\frac{2 \gamma  \delta a}{f_I \theta_i} ~, \nonumber \\
  \gamma&\equiv& \frac{\Omega_a }{\Omega_d }~,\nonumber\\
  \Omega_a &\simeq& 0.27\times \theta_i^2\bigg(\frac{f_a}{10^{12}~\text{GeV}}\bigg)^\frac{7}{6}~,
\end{eqnarray}
where assuming $\delta \theta \ll \theta_i$, we obtain the axion isocurvature amplitude $S_a \approx 2 \delta\theta/\theta_i$. The dark matter isocurvature, $S_d$ is a product of $S_a$ and $\gamma$, the fraction of axion DM relic abundance $\Omega_a h^2$ among the total DM abundance $\Omega_d h^2$. Throughout the article, we work in the framework of single field inflation and assume that the axion isocurvature is the only source of DM isocurvature. We do not consider the more complicated multifield inflation scenario in which DM isocurvature could be generated during reheating of multiple inflatons~\cite{Martin:2021frd}. Consequently, the remaining part of DM are assumed to be generated from radiation and will not contribute to the DM isocurvature $S_d$. The expression for $\Omega_a$ in the last line, adapted from~\cite{Dine:2017swf}, is for QCD axion through the vanilla misalignment mechanism~\cite{Preskill:1982cy,Dine:1982ah,Abbott:1982af}. Note that $\Omega_a$ depends on $f_a$, the axion decay constant after inflation.\footnote{Beyond the minimal misalignment mechanism, mechanisms such as early matter domination~\cite{Lazarides:1990xp,Kawasaki:1995vt, Banks:1996ea, Giudice:2000ex,Grin:2007yg}, thermal friction from gauge fields~\cite{Berghaus:2019cls,Papageorgiou:2022prc,Choi:2022nlt}, or dynamical PQ scale~\cite{Allali:2022sfm,Allali:2022yvx} could significantly alternate $\Omega_a$, which we will not explore further in the paper. Alternatively, an axion-like particle could also go through the misalignment mechanism and become DM. The axion-like particle's mass is not fixed by the standard model QCD scale, introducing an extra free parameter when calculating its relic density. In all scenarios above, the corresponding decay constants could be significantly higher than in the vanilla case. Elaborate discussions for these possibilities will be left to future studies.} As we will discuss in more detail, $f_I$ is not necessarily equal to $f_a$ in the presence of the PQ field coupling to the inflaton. 

The isocurvature spectrum is then given by 
\begin{equation}
    P_i = A_i \left(\frac{k}{k_0}\right)^{1-n_i}~,
\end{equation}
where the amplitude is given by 
\begin{equation}
    A_i = \left(\frac{\gamma H}{\pi f_I \theta_i}\right)^2~,
\end{equation}
and $n_i$ is the isocurvature spectral index. 
The current CMB measurements constrain uncorrelated DM isocurvature to be~\cite{Planck:2018jri} 
\begin{equation}
    \beta \equiv \frac{A_i}{A_s} = \frac{1}{A_s} \left(\frac{\gamma H}{\pi f_I \theta_i}\right)^2 < 0.038~, 
    \label{eq:isoconstraints}
\end{equation}
where $A_s \equiv H^4/((2 \pi)^2 \dot{\phi}_0^2) \approx 2.1 \times 10^{-9}$ is the curvature power spectrum amplitude~\cite{Planck:2018jri}. As we will show in the remaining sections, to get interesting signals in our models at the CMB scales, we focus on the parameter space with $\gamma \ll 1, \theta_i \sim 1$, and $f_I/H \sim {\cal O}(10 -100)$.\footnote{A $\theta_i$ of $\mathcal{O}(1)$ is a natural expectation if the initial phase is random drawn from a uniform distribution between $(-\pi, \pi]$. The discussions can be extended to smaller $\theta_i$ values until $\theta_i$ becomes comparable to the tiny $\delta \theta$. A more precise expansion in Eq.~\eqref{eq:isocurvatureanda} will be needed in this case. However, as long as $S_{a}$ is still related to $\delta a$, the isocurvature feature will still appear.}

\section{Models and effective operators}
\label{sec:operators}

We consider the following PQ field model during inflation with the $(-, +, +, +)$ signature:
\begin{align}
 {\cal L} = -\frac{(\partial_\mu \phi)^2}{2} - |\partial_\mu \chi|^2 - V_\phi(\phi)- V_\chi(\chi) - V_{\rm add}~,~
	 V_\chi(\chi)  =   \frac{\lambda}{2} \bigg(|\chi|^2 - \frac{f_a^2}{2} \bigg)^2~, \label{eqn:masterLagrangian}	
\end{align}
where $\phi$ is the inflaton with a potential $V_\phi(\phi)$, $\chi$ is the complex PQ scalar field with a potential $V_\chi(\chi)$, and $f_a$ is the axion decay constant or equivalently the VEV of the PQ scalar {\it today}. $V_{\rm add}$ includes the leading-order high-dimensional operators coupling the PQ field to the inflaton or other heavy spectator fields (e.g., we will consider some heavy fermions charged under $U(1)_{\rm PQ}$, $\psi$) during inflation, which could lead to different types of interesting spectrum or CC signals. We require these operators to respect the (approximate) shift symmetry of the inflaton and the PQ symmetry $U(1)_{\rm PQ}$. The operators and their corresponding cosmological signals in density perturbations are listed in Table~\ref{tab:models}. 

\begin{table}[h!]
\begin{center}
\begin{tabular}{| c|l | l |l |} 
 \hline
Model & New cosmological signals & $V_{\rm add}$  & Section \\
\hline
1   & \color{purple}{Correlated clock signals} & $\frac{c}{\Lambda^2} (\partial_\mu \phi)^2\,|\chi|^2$ & \ref{ssec:model1twopoint} \\
& \color{purple}{in curvature and isocurvature power spectra;} &  &\\
& \color{purple}{Mixed isocurvature-curvature bispectrum; } &  &\ref{ssec:model1threepoint}\\
\hline 
2 & \color{purple}{Clock signal in isocurvature power spectrum;} & $\frac{\tilde{c}}{\Lambda}\partial_\mu \phi\,\partial^\mu(|\chi|^2)$ &\ref{ssec:model2twopoint} \\
 & \color{purple}{Mixed isocurvature-curvature bispectrum;} & &\ref{ssec:model2threepoint} \\
 \hline
3  &\color{orange}{Correlated mixed and pure isocurvature} &$i\frac{\kappa \partial_\mu \phi}{\Lambda} (\chi^\dagger \partial^\mu \chi - \chi \partial^\mu \chi^\dagger)$  & \ref{sec:model3}  \\
&\color{orange}{bispectra.} & $+ y \chi \psi_L^\dagger \psi_R + c.c.$  & \\
\hline
\end{tabular}
\end{center}
\caption{Three simple PQ scalar models that lead to novel inflationary signals which involves isocurvature fluctuations. The signals in {\color{purple}{purple}} font indicate the signals enhanced due to sharp features in the inflationary potential, while those in {\color{orange}{orange}} indicate the signals enhanced due to chemical-potential type couplings.  }
\label{tab:models}
\end{table}

In each model in Table~\ref{tab:models}, we have only one high-dimensional operator coupling the PQ field to the inflaton for simplicity. In principle, one could combine multiple operators in one model, which may lead to more complicated new CC signals. Note that $V_{\rm add}$ in model 1 is a dimension-6 operator, in contrast to the dimension-5 operators in the other two models. The reason for this slightly odd ordering is that model 1 could lead to a striking signal, the correlated clock signals in the curvature and isocurvature power spectra, which contain most information and could probably be most easily tested with future cosmological observations. Also, although more suppressed by the cutoff scale $\Lambda$, the amplitude of signal in model 1 is not necessarily more suppressed than that in model 2 due to different dependence on other parameters.

In all the models, we focus on the scenario in which the PQ symmetry is spontaneously broken so that $\chi$ acquires a VEV, $f_I$, during inflation. As we will show, $f_I$ could be shifted from $f_a$ in the presence of $V_{\rm add}$. Then $\chi$ could be parameterized as 
\begin{equation}
\chi = \frac{f_I + \sigma}{\sqrt{2}} e^{i a/f_I}~, 
\label{eq:chiparametrization}
\end{equation}
where $\sigma$ is the heavy radial mode and the massless phase mode $a$ serves as the axion, which could become (part of) the cold DM after inflation. As will be discussed in the models, we expect that the radial mode mass, about the same size of $f_I$, is much above the inflationary Hubble scale. 

The amplitude of cosmological production of heavy particles, which is $\sigma$ in our case, is proportional to $e^{-\pi m/H}$ with $m$ the mass of the heavy particle (see also \cite{Li:2019ves,Sou:2021juh}). Thus the corresponding CC signal gets exponentially suppressed when $m \gg H$. This clearly challenges the power of CC probe to super-$H$ scale and the inflationary axion scenario. To overcome this difficulty, we consider two possible enhancement mechanisms which have already been proposed in the CC literature. The first one is to employ the primordial features, some shift-symmetry violating features along the inflaton trajectory in the inflationary landscape. These features could excite a heavy field with $m \gg H$ classically and quantum-mechanically, resulting in sizable signals in the power spectrum and non-Gaussianities~\cite{Chen:2011zf,Chen:2022vzh}. The other one is to use a chemical-potential type coupling $\partial_\mu \phi J^\mu$, with $J^\mu$ a current made of the heavy fields~\cite{Chen:2018xck,Hook:2019zxa,Hook:2019vcn,Wang:2019gbi,Wang:2020ioa,Bodas:2020yho,Sou:2021juh}. This coupling still preserves the inflaton shift symmetry but can introduce a new scale $\dot{\phi}_0/\Lambda \gg H$ when the inflaton is set to its homogeneous background value~\cite{Chen:2018xck,Hook:2019zxa,Hook:2019vcn,Wang:2019gbi,Wang:2020ioa,Bodas:2020yho,Sou:2021juh}. This new higher scale allows excitation and production of heavy particles, compensating the Boltzmann suppression with $e^{\pi\dot{\phi}_0/(\Lambda H)}$ and boosting the NG signals. In our model 1 and 2, we will rely on the primordial feature mechanism, while in model 3 we employ the chemical-potential coupling.

\section{Model 1: $(\partial \phi)^2|\chi|^2$}
\label{sec:model1}

The Lagrangian of our first model is given by 
\begin{align}
 {\cal L}_1 &= -\frac{(\partial_\mu \phi)^2}{2} - |\partial_\mu \chi|^2 - V_\phi(\phi)- V_\chi(\chi) - \frac{c}{\Lambda^2} (\partial \phi)^2 |\chi|^2~, \nonumber \\
 &= -\frac{(\partial_\mu \phi)^2}{2} - \frac{(\partial_\mu \sigma)^2}{2}- \frac{1}{2}\left(1+\frac{\sigma}{f_I}\right)^2(\partial_\mu a)^2 - V_\phi(\phi)- V_\sigma(\sigma) - \frac{c}{2\Lambda^2}(\partial_\mu \phi)^2 \left(f_I + \sigma\right)^2~,
\label{eqn:Lagrangian2}
\end{align}
where $c$ is a positive dimensionless coefficient and $\Lambda$ is the cutoff energy scale of the model.\footnote{The $c$ coefficient could be negative as well. For a negative $c$, small $|c|$ will suppress the signals we will discuss in the rest of the paper, as the sizes of the signals are proportional to powers of $|c|$. Large $|c|$ could restore the PQ symmetry during inflation and no isocurvature mode will be generated~\cite{Bao:2022hsg}. We will not discuss these possibilities further.} Throughout this work, we assume such model parameters to be constants independent of $\phi$ for simplicity. In the second line above, we use the parametrization in Eq.~\eqref{eq:chiparametrization} after the PQ symmetry breaking, and $V_\sigma(\sigma)$ is given by
\begin{equation}
V_\sigma(\sigma) =  \frac{\lambda}{8} \bigg(\left(f_I + \sigma\right)^2 - f_a^2 \bigg)^2~.
\label{eq:Vrho}
\end{equation}
Note that in this model, the axion field does not couple directly to the inflaton.
The VEV of the PQ field during inflation is given by 
\begin{equation}
f_I^2  = f_a^2  +\frac{2c \dot{\phi}_0^2}{\lambda \Lambda^2}~,
\label{eq:fIfa}
\end{equation}
where $\dot{\phi}_0$ is the homogeneous background of the inflaton (the standard attractor solution), satisfying
\begin{equation}
\dot{\phi}_0 \approx -\frac{1}{3H} \frac{\partial V_\phi}{\partial \phi} = \sqrt{\frac{2}{3} \epsilon V_\phi}~.
\label{eq:slowroll}
\end{equation}
The equation above assumes the slow-roll approximation with $\left|\ddot{\phi}\right| \ll \left|\frac{\partial V_\phi}{\partial \phi}\right|$, and the slow-roll parameter is $\epsilon=M_{\rm pl}^2 (\partial_\phi V_\phi/V_\phi)^2/2$ with $M_{\rm pl}\approx 2.4 \times 10^{18}$ GeV the reduced Planck scale. In single field inflation, $\dot{\phi}_0 \approx (60 H)^2$, fixed by the amplitude of the density perturbations. 
The mass squared of the radial mode is 
\begin{equation}
m_{\sigma, \rm eff}^2 = \lambda f_I^2~.
\label{eq:mrhofI}
\end{equation}

For the model to be a valid effective field theory (EFT), we need to impose the following constraints following approaches in~\cite{Kumar:2017ecc,Bao:2022hsg}:
\begin{enumerate}[label=(\alph*)]
\item The added inflaton-PQ field coupling in this model is just the leading one of a series $(\partial\phi)^{2n} |\chi|^2/\Lambda^{4n-2}$. Requiring this power expansion doesn't spoil the EFT during inflation, we need to impose~\cite{Creminelli:2003iq}
\begin{align}
\Lambda > \sqrt{\dot{\phi}_0}~.
\end{align}
\item Naturalness constraint on the quantum correction to $\lambda$ from the inflaton loop with the cutoff $\sim \Lambda$: 
\begin{align}
c \lesssim 4 \pi \sqrt{\lambda}~.
\end{align}
\item Naturalness constraint on the quantum correction to $m_{\sigma, \rm eff}^2$ from the inflaton loop with the cutoff $\sim \Lambda$: 
\begin{align}
\frac{c \Lambda^2}{16 \pi^2} \lesssim m_{\sigma, \rm eff}^2 =  \lambda f_I^2~. 
\end{align}
\item The high-dimensional operator only leads to a small correction to the inflaton kinetic term of $(\partial\phi)^2$: 
\begin{align}
q \equiv \frac{c f_I^2}{\Lambda^2} \ll 1~.
\label{eq:kinetic}
\end{align}
\end{enumerate}

\begin{figure}[h!]
    \centering
    \includegraphics[width=7.5cm]{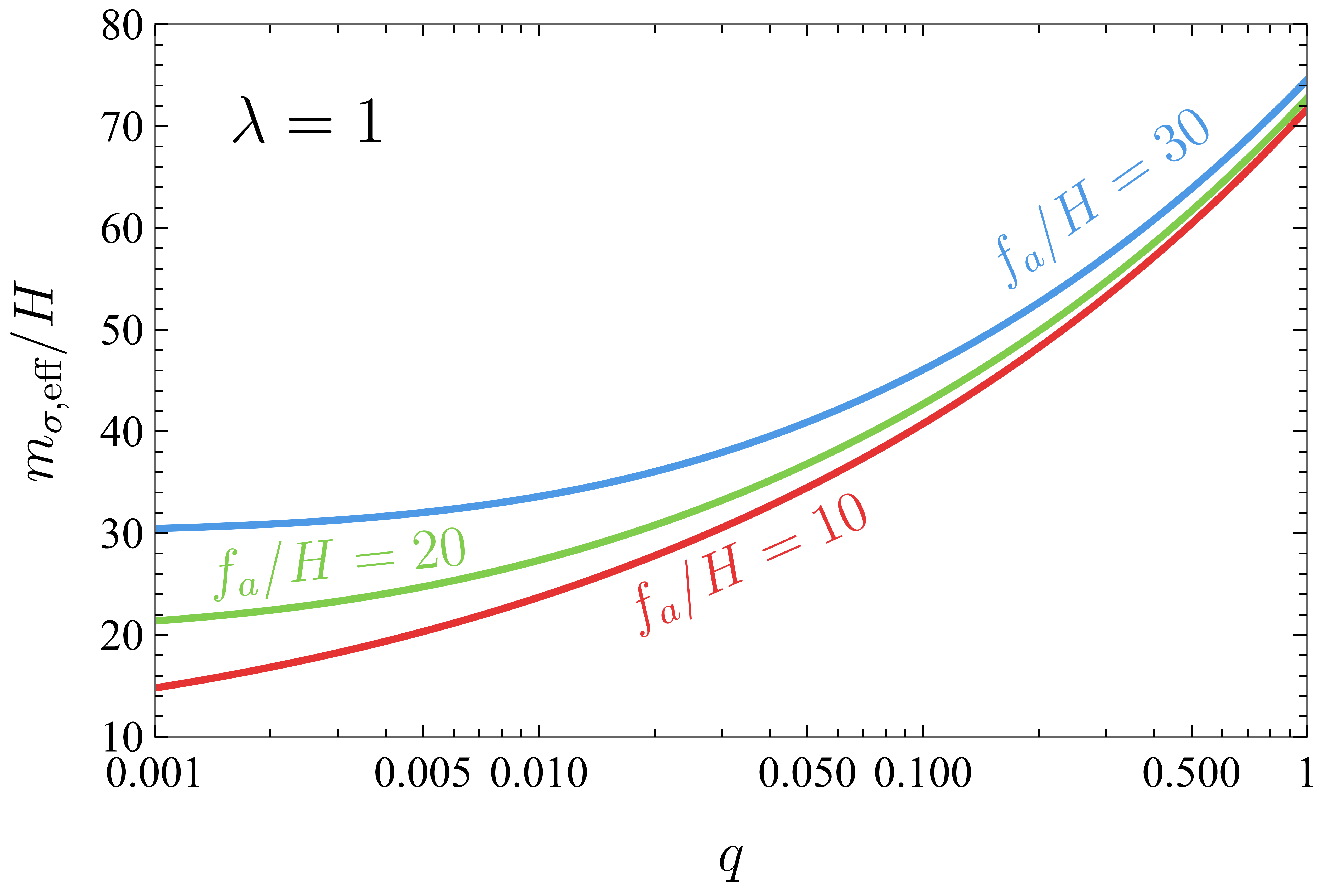}  \includegraphics[width=7.5cm]{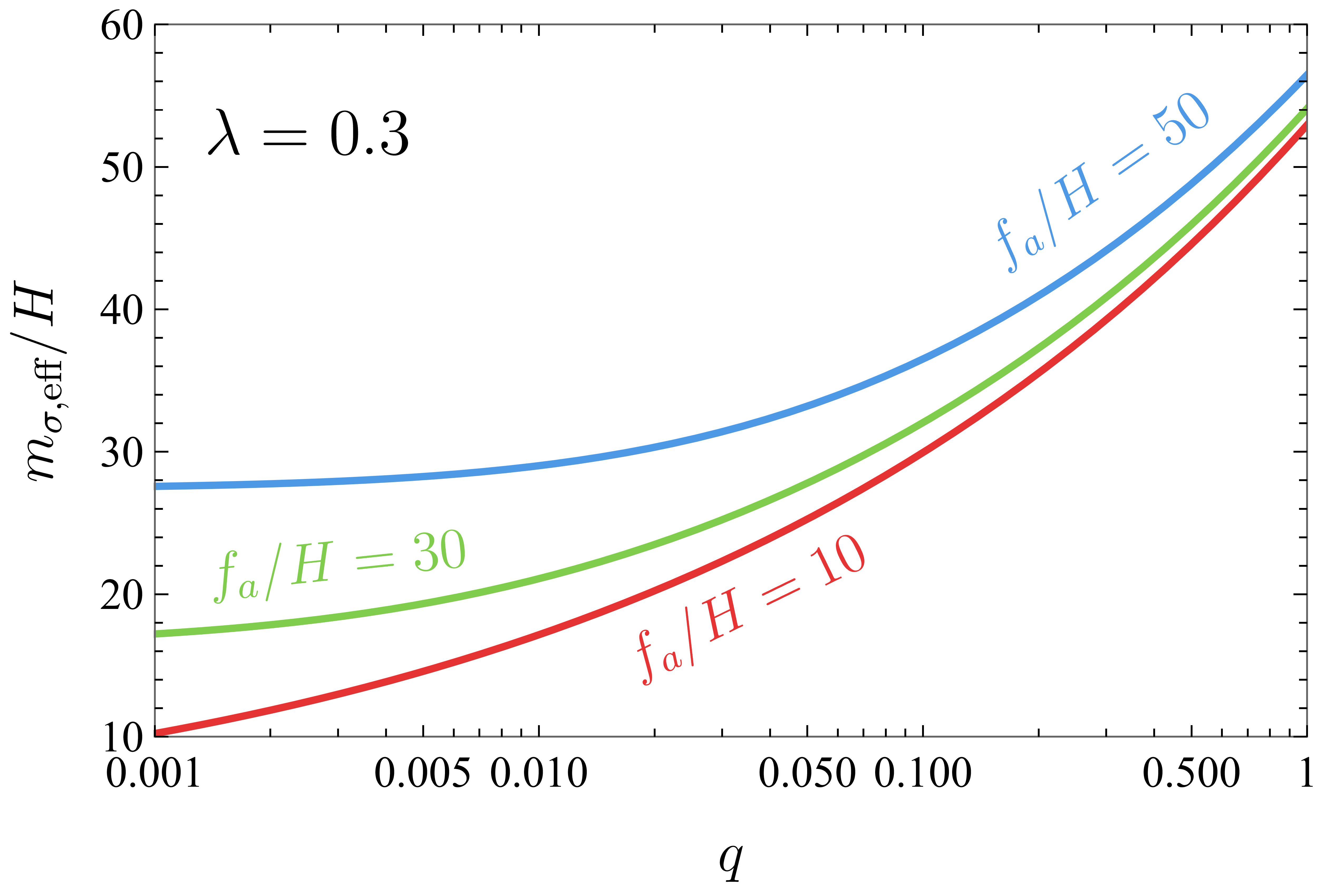}
    \caption{$m_{\sigma, \rm eff}/H$ as a function of $q$ for different choices of $f_a/H$. Left: $\lambda=1$; right: $\lambda = 0.3$. }
    \label{fig:fIHq}
\end{figure}

Eq.~\eqref{eq:fIfa} and Eq.~\eqref{eq:mrhofI} allow us to solve $f_I/H$ and $m_{\sigma, \rm eff}/H$ as a function of $f_a/H, q$ and $\lambda$. For the natural value of $\lambda \sim 1$  and different choices of $f_a/H \subset {\cal O}(10)$, we show $m_{\sigma, \rm eff}/H$ as a function of $q$ in Fig.~\ref{fig:fIHq}. 
One can see that when $q$ increases, the contribution to $m_{\sigma, \rm eff}/H$ from the inflaton-PQ field coupling dominates over that from the PQ-field potential, approaching an asymptotic value $(2\lambda)^{1/4}\sqrt{\dot{\phi}_0}/H$ at $q=1$. Smaller $\lambda$ allows for larger $f_a/H$ to achieve the same $m_{\sigma, \rm eff}/H$. 

In all the models we consider, $m_{\sigma, \rm eff} = \sqrt{\lambda} f_I \gg H$ during inflation, the cosmological production of $\sigma$, and the induced CC signals will be exponentially suppressed by $\sim e^{-\pi m_{\sigma, \rm eff}/H}$, as mentioned in the summary of Sec.~\ref{sec:operators}. To enhance the signals, we need to introduce some boosting mechanisms. In model 1 and 2, we will rely on the primordial feature mechanism in which the inflaton potential possesses a sharp feature and can be decomposed as 
\begin{equation}
V_\phi = V_{\phi 0} + V_{\phi 1}~, ~~|V_{\phi 1}| \ll |V_{\phi 0}|~,
\end{equation}
where $V_{\phi 0}$ is the smooth potential responsible for the featureless attractor solution in Eq.~\eqref{eq:slowroll}, while the perturbation $V_{\phi 1}$ is a small sudden change of the potential localized in some regions of the field space. Then the classical background trajectories can be split into the featureless (with subscript $0$) and featured (with subscript $1$) components correspondingly:
\begin{equation}
\phi_{\rm bkg}(t) = \phi_0(t)+ \phi_1(t)~,~~
\sigma_{\rm bkg} (t) = \sigma_0(t)+\sigma_1(t) = \sigma_1(t)~,
\end{equation}
where $\phi_0(t)$ is given in Eq.~\eqref{eq:slowroll} and $\sigma_0(t)=0$. Note that since $a$ does not couple to the inflaton directly, $a_{\rm bkg}(t)$ is a constant.

There are many examples of sharp features. Such features can be easily envisioned when we embed the inflation models in a potential landscape when the universe is unstable. There are also motivations from CMB data analyses \cite{Peiris:2003ff,Adams:2001vc,Bean:2008na, Mortonson:2009qv, Hazra:2010ve, Hazra:2014goa, Miranda:2014fwa, Braglia:2021sun, Braglia:2021rej, Braglia:2022ftm}. A sharp feature could naturally excite the classical oscillation of a massive field, such as the radial mode of the PQ field, which is otherwise hard to probe because its mass is much larger than the Hubble scale.
In this paper, as a toy example, we consider $V_{\phi 1}$ to be a sharp step function: 
\begin{align}
V_{\phi1}(\phi) &= - b  V_{\phi 0} \, \theta(\phi-\phi_s)~, \label{eq:Vphi11} 
\end{align}
where $b \subset (0,1)$ is a dimensionless quantity; $\theta(\phi-\phi_s)$ is the Heaviside $\theta$ function: it is one when $\phi > \phi_s$ and zero otherwise, which means that the perturbation potential is a small but sharp downward-step. Similar or other examples of sharp features with a more complicated form have been used to explain the CMB residual anomalies~\cite{Peiris:2003ff,Adams:2001vc, Bean:2008na, Mortonson:2009qv, Hazra:2010ve, Hazra:2014goa, Miranda:2014fwa, Braglia:2021sun, Braglia:2021rej, Braglia:2022ftm}. Here we just use this simplest form to illustrate the new clock and CC observables. 

\begin{figure}[h!]
    \centering
    \includegraphics[width=\textwidth]{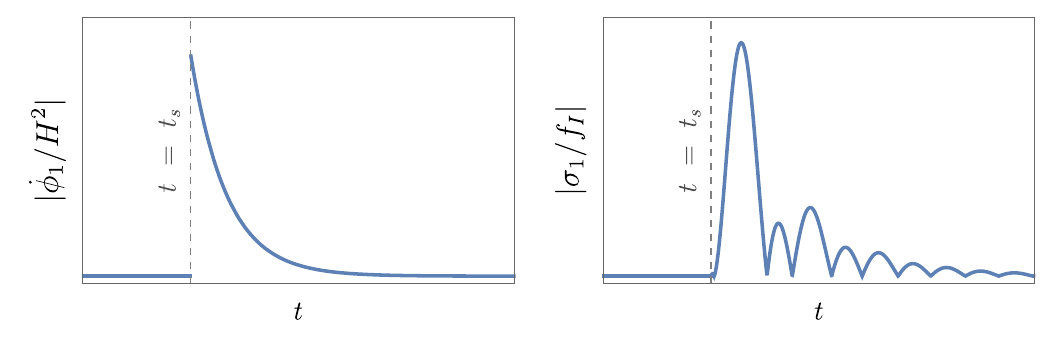}
    \caption{The scenario of a sharp feature given Eq.~\eqref{eq:Vphi11}. Left: $\left|\dot{\phi}_1/H^2\right|$ as a function of $t$; right: induced radial mode oscillations $|\sigma_1/f_I|$ as a funtion of $t$. }
    \label{fig:step}
\end{figure}

The time evolution of $\phi_1$ and $\sigma_1$ due to $V_{\phi1}$ follow the equations of motion (EOMs) below: 
\begin{align}
   & \ddot{\phi}_1 +3 H\dot{\phi}_1 + \left.\frac{\partial{V_{\phi 1}}(\phi)}{\partial \phi}\right|_{\phi=\phi_0}  \simeq 0~, \\
 &\ddot{\sigma}_1 +3 H\dot{\sigma}_1 + \frac{c f_I}{\Lambda^2} \left(\dot{\phi}_{\rm bkg}^2 - \dot{\phi}_0^2\right)  +m_{\sigma;\rm eff}^2 \sigma_1 \simeq 0~ \nonumber \\
  \Rightarrow~&\ddot{\sigma}_1 +3 H\dot{\sigma}_1 + \frac{c f_I}{\Lambda^2} \left(2 \dot{\phi}_0 \dot{\phi}_1\right)  +m_{\sigma;\rm eff}^2 \sigma_1 \simeq 0~,
\end{align}
where we approximate $\dot{\phi}_{\rm bkg}^2 - \dot{\phi}_0^2\approx 2 \dot{\phi}_0 \dot{\phi}_1$. Given Eq.~\eqref{eq:Vphi11}, the solutions are 
\begin{align}
\phi_1(t) &= \frac{b  V_{\phi 0}}{3 H \dot{\phi}_0} \left[1-e^{-3H(t-t_s)}\right]\theta(t-t_s)~, \label{eq:dim5featuresphi1} \\
\sigma_1(t) &\simeq \frac{2c \, b V_{\phi 0} f_I}{\Lambda^2\, m^2_{\sigma,{\rm eff}}}\left\{e^{-\frac{3}{2}H(t-t_s)} \cos \left[\mu_\sigma H (t-t_s)\right]  -e^{-3H(t-t_s)}\right\} \theta(t-t_s)~, \label{eq:dim5featuresrho1} \nonumber \\
 \mu_\sigma &\equiv \sqrt{\frac{m^2_{\sigma,{\rm eff}}}{H^2} -\frac{9}{4}}~,
\end{align}
where $t_s$ corresponds to the time when the inflaton passes through $\phi_s$ and we ignore higher-order terms suppressed by more powers of $m_{\sigma,{\rm eff}}/H \simeq \mu_\sigma$ in $\sigma_1(t)$. An illustration of this scenario is presented in Fig.~\ref{fig:step}. The inflaton velocity has a steep jump at $t=t_s$ with $(\dot\phi_{\rm bkg}^2 - \dot\phi_0^2)/\dot{\phi}_0^2|_{t=t_s} \approx {2b  V_{\phi 0}}/{\dot{\phi}_0^2}$. 
This triggers oscillation in the radial mode $\sigma$ with an amplitude proportional to $c$ and suppressed by $m^2_{\sigma, {\rm eff}}$ or equivalently $\mu_\sigma^2$ when $\mu_\sigma \gg 1$. 

\subsection{Corrections to the power spectra}
\label{ssec:model1twopoint}
Now we move on to consider the perturbations, defined as $\delta \phi= \phi - \phi_{\rm bkg}$, $\delta \sigma = \sigma -\sigma_{\rm bkg}$, and $\delta a = a - a_{\rm bkg}$, and compute their two-point correlators.
Expanding Eq.~\eqref{eqn:Lagrangian2}, we have the following leading quadratic terms
\begin{align}
  {\cal L}_{1}^{(2)}  & \supset - \frac{1}{2}\frac{\partial^2 V_\phi}{\partial \phi^2} (\delta \phi)^2 \nonumber \\
    &+ \left( \frac{1}{2} + \frac{\sigma_{\rm bkg}}{f_I}\right) \left((\delta \dot{a})^2 - \frac{1}{R^2}(\partial_i \delta a)^2\right)\nonumber \\
    &+ \frac{c f_I^2}{\Lambda^2} \left( \frac{1}{2} + \frac{\sigma_{\rm bkg}}{f_I}\right)  \left(({\delta \dot{\phi}})^2- \frac{1}{R^2}(\partial_i \delta \phi)^2\right) \nonumber \\
    &+ \frac{2c f_I \dot{\phi}_0}{\Lambda^2} \left(1+ \frac{\dot{\phi}_1}{\dot{\phi}_0}+\frac{\sigma_{\rm bkg}}{f_I}\right) \delta \dot{\phi} \delta \sigma~,
    \label{eq:model1quadratic}
\end{align}
where $R$ is the scale factor; the overhead dot indicates the derivative with respect to time while $\partial_i$ is the derivative of the spatial coordinate. We ignore the contribution from the inflaton kinetic term, which does not contribute to the power spectra corrections. We also ignore terms of higher order ${\cal O}((\sigma_{\rm bkg}/f_I)^2)$ that are highly suppressed by small parameters. Finally, terms proportional to $\delta\sigma^2$ are dropped as they only contribute to the two-point correlators at higher order in $c/\Lambda^2$; the $\delta \dot{\phi} \delta \sigma$ term also contributes to two-point correlators at higher order, but we include it here since they will contribute to the three-point functions which we will discuss later.

\noindent {\underline {\it Sharp feature signal in curvature power spectrum}} We first compute the leading correction to the curvature power spectrum due to the step function potential in Eq.~\eqref{eq:Vphi11}, through the first term in Eq.~\eqref{eq:model1quadratic}:
\begin{equation}
\frac{\partial^2 V_{\phi}}{\partial \phi^2} (\delta \phi)^2 \simeq \frac{\partial^2 V_{\phi1}}{\partial \phi^2} (\delta \phi)^2 =-\frac{b V_{\phi0} H^2 \tau}{\dot{\phi}_0^2}  \tau_s \;\left(\partial_\tau\delta (\tau-\tau_s)\right)(\delta \phi)^2~,
\label{eq:Vpp}
\end{equation}
where $\tau$ is the conformal time with $d\tau = dt/R$ and $R(\tau)=-1/(H \tau)$. $\tau_s$ is the corresponding conformal time at the transition point $\phi_s$. 

\begin{figure}[h!]
    \centering
    \includegraphics[width=7cm]{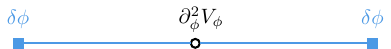}
    \caption{Diagram of the sharp feature signal in the curvature power spectrum. The external points of inflaton are marked with blue squares. The solid blue line represents the propagator for $\phi$ while the empty dot denotes the interaction in Eq.~\eqref{eq:Vpp}.  }
    \label{fig:curvature2pt1}
\end{figure}

An insertion of this interaction leads to the diagram in Fig.~\ref{fig:curvature2pt1}, resulting in a correction to the curvature power spectrum $P_\zeta$ as 
\begin{align}
\left. \quad\frac{\Delta P_\zeta}{P_\zeta}\right|_{\rm sharp} & = -\bigg(\frac{-1}{2} \bigg)(-2 i) \int_{-\infty}^0 \frac{d\tau}{(H\tau)^4} u_k^2 \frac{b V_{\phi 0}H^2 \tau \tau_s}{\dot{\phi}_0^2}  \;\left[\partial_\tau\delta (\tau-\tau_s)\right] + \text{c.c.}~,\nonumber \\
    &= - \frac{b V_{\phi 0}}{\dot{\phi}_0^2} g\left(\frac{k}{k_0}\right) \sin\bigg(\frac{2k}{k_0}+ \omega \bigg)~, \label{eq:correctioncur2pt1}
\end{align}
where $u_k=\frac{H}{\sqrt{2 k^3}}(1 + i k \tau) e^{-i k \tau}$ is the wavefunction of the massless inflaton mode; $k_0 =-\tau_s^{-1}$ is the reference wavenumber for the sinusoidal signal; the function $g(k/k_0)$ determines the envelope of the signal and $\omega$ gives the phase. This sinusoidal correction is entirely due to the step feature in time and is independent of the excited oscillating heavy $\sigma$ field. Thus such a signal does not contain any information about the PQ sector and its coupling to the inflaton. 
Both $g(k/k_0)$ and $\omega$ are model-dependent, sensitive to the input perturbation potential $V_{\phi1}$. Their detailed forms are not important for our following discussions, so we will not list them here. Instead we will focus on the amplitude at $k\sim k_0$, which is of order $\sim b V_{\phi0}/\dot{\phi}_0^2$ [$g(k/k_0) \sim {\cal O}(1)$ when $k \sim k_0$].   
 There is a suggestive dip feature in the power spectrum present near $\ell \sim 25$ in the CMB, which could be interpreted as a sharp feature signal with $\Delta P_\zeta/P_\zeta\sim 0.3$~\cite{Peiris:2003ff, Braglia:2021rej}. Taking this as a rough guideline, we consider $b V_{\phi0}/\dot{\phi}_0^2 \lesssim 0.3$ for evaluating the more important and interesting clock signals below.

The abrupt starting point of the oscillation in $\sigma_{\rm bkg}$ at $t=t_s$ is also a potential source of sharp features. Due to the couplings in the 2nd and 3rd lines of \eqref{eq:model1quadratic}, this sharp feature contributes to additional sharp feature signals in both curvature and isocurvature power spectra. We will not present any details of such signals in this paper, only to mention that common to all sharp feature signals, these signals would also have the sinusoidal-running scale dependence, and model-dependent envelops. The amplitudes of these signals would be smaller than the ones we considered above, due to the sizes of the couplings and the milder nature of the sharp feature.

\begin{figure}[h!]
    \centering
    \includegraphics[width=\textwidth]{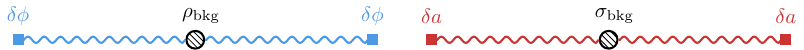}
    \caption{Diagrams of the clock signals in the curvature (left) and isocurvature (right) power spectra. The external points of the axion (inflaton) are marked with red (blue) squares. The solid red (blue) line represents the propagator for $a$ ($\phi$) while the shaded blob denotes the interaction in either the second or the third line of Eq.~\eqref{eq:model1quadratic}, with an insertion of the oscillating $\sigma_{\rm bkg}$.  }
    \label{fig:clocksignal}
\end{figure}

\noindent {\underline {\it Correlated clock signals in the power spectra}} 
Now we consider the corrections to either the curvature or isocurvature power spectra due to the resonances between the induced oscillating heavy radial mode and the axion or inflaton, which result in clock signals with rich information for the PQ field and its couplings during inflation. The leading diagrams, from interactions in the second and third lines of Eq.~\eqref{eq:model1quadratic} with insertions of $\sigma_{\rm bkg}$, are presented in Fig.~\ref{fig:clocksignal}.

We take $\sigma_{\rm bkg}$ to be Eq.~\eqref{eq:dim5featuresrho1} and compute the two diagrams in the same way:
\begin{align}
 \left. \frac{\Delta P_\zeta}{P_\zeta} \right|_{\rm clock} &= \frac{-2 i c f_I}{\Lambda^2} \int_{-\infty}^0 \frac{d\tau}{(H\tau)^4} \bigg( \dot{u}_k^2 - \frac{k^2}{R^2} u_k^2 \bigg)\sigma_{\rm bkg} + \text{c.c.}~,\nonumber \\
&\simeq -\frac{ 2c^2  b V_{\phi 0} f_I^2}{\Lambda^4 m^2_{\sigma,{\rm eff}}} \sqrt{2\pi\mu_\sigma}  \left(\frac{k}{k_r}\right)^{-3/2} \sin\left[\mu_\sigma \ln\left(\frac{k}{k_r}\right) + \mu_\sigma + \frac{\pi}{4}\right]~, \label{eq:clockcur}  \\
 \left. \frac{\Delta P_i}{P_i} \right|_{\rm clock} &\simeq   \left.\frac{\Delta P_a}{P_a}\right|_{\rm clock}= \frac{-2 i}{f_I} \int_{-\infty}^0 \frac{d\tau}{(H\tau)^4} \bigg( \dot{u}_k^2 - \frac{k^2}{R^2} u_k^2 \bigg)\sigma_{\rm bkg} + \text{c.c.}~,\nonumber \\
&\simeq - \frac{2 c  b V_{\phi 0}}{\Lambda^2 m^2_{\sigma,{\rm eff}}} \sqrt{2\pi\mu_\sigma} \left(\frac{k}{k_r}\right)^{-3/2} \sin\left[\mu_\sigma \ln\left(\frac{k}{k_r}\right) + \mu_\sigma + \frac{\pi}{4}\right]~, \label{eq:clockiso}  
\end{align}
where $k_r = -\mu_\sigma/ (2\tau_s)=\mu_\sigma k_0/2$. 
We remind that the clock signals are generated through the resonance between the harmonic (standard) oscillation of the massive field and the oscillation of the inflaton or axion quantum fluctuation mode \cite{Chen:2011zf,Chen:2011tu}. Because the time dependence of the latter frequency is determined by the background scale factor evolution $a(t)$, the momentum dependence in the phase of the clock signal takes the form of the inverse function of $a(t)$, which in this case is a logarithmic function.
The $\sqrt{2\pi\mu_\sigma}$ enhancement arises from the resonance \cite{Chen:2008wn,Flauger:2009ab,Flauger:2010ja,Chen:2010bka}.
Note that $\left|\frac{\Delta P_i}{P_i} \right|_{\rm clock}$ is equal to the oscillating part of $|\sigma_1/f_I|$ times $\sqrt{2 \pi \mu_\sigma}$. In other words, $|\sigma_1/f_I|$ is suppressed by a factor of $\sqrt{2 \pi \mu_\sigma} \sim {\cal O}(10)$ compared to $\left|\frac{\Delta P_i}{P_i} \right|_{\rm clock}$. The two corrections, \eqref{eq:clockcur} and \eqref{eq:clockiso},  are the same up to an overall rescaling factor $c f_I^2/\Lambda^2$, which originates from the different coefficients of the quadratic terms for $(\delta a)^2$ and $(\delta \phi)^2$ in Eq.~\eqref{eq:model1quadratic}. 

Let's estimate the amplitudes of the corrections at $k=k_r$. In the limit $\mu_\sigma \gg 1$,
\begin{align}
\left| \frac{\Delta P_\zeta}{P_\zeta}\right|_{\rm clock; amp} &= \frac{2c^{2} b V_{\phi 0}f_I^2}{\Lambda^4 H^2} \sqrt{\frac{2\pi }{\mu_\sigma^3}} \nonumber \\
  &\approx ~ 0.019 \left(\frac{q}{0.02}\right)^2 \left(\frac{b V_{\phi 0}}{0.3 \dot{\phi}_0^2} \right) \left(\frac{\dot{\phi}_0}{(60H)^2}\right)^2 \left(\frac{40 H}{f_I}\right)^{7/2} \left(\frac{1}{\lambda}\right)^{3/4}~, \label{eq:model1Pcclock} \\
  \left| \frac{\Delta P_i}{P_i}\right|_{\rm clock;amp} &\approx \frac{2c b V_{\phi 0}}{\Lambda^2 H^2} \sqrt{\frac{2\pi }{\mu_\sigma^3}} \nonumber \\
 &\approx ~ 0.96 \left(\frac{q}{0.02}\right) \left(\frac{b V_{\phi0}}{0.3 \dot{\phi}_0^2} \right) \left(\frac{\dot{\phi}_0}{(60H)^2}\right)^2 \left(\frac{40 H}{f_I}\right)^{7/2} \left(\frac{1}{\lambda}\right)^{3/4}, \label{eq:model1Piclock}
\end{align}
where we use $\mu_\sigma \approx m_{\sigma;\rm eff}/H = \sqrt{\lambda} f_I/H$ and $q \equiv c f_I^2 /\Lambda^2 \ll 1$ to satisfy the constraint on the kinetic term of $\phi$ as discussed around Eq.~\eqref{eq:kinetic}.

\begin{figure}[h!]
    \centering
    \includegraphics[width=7.5 cm]{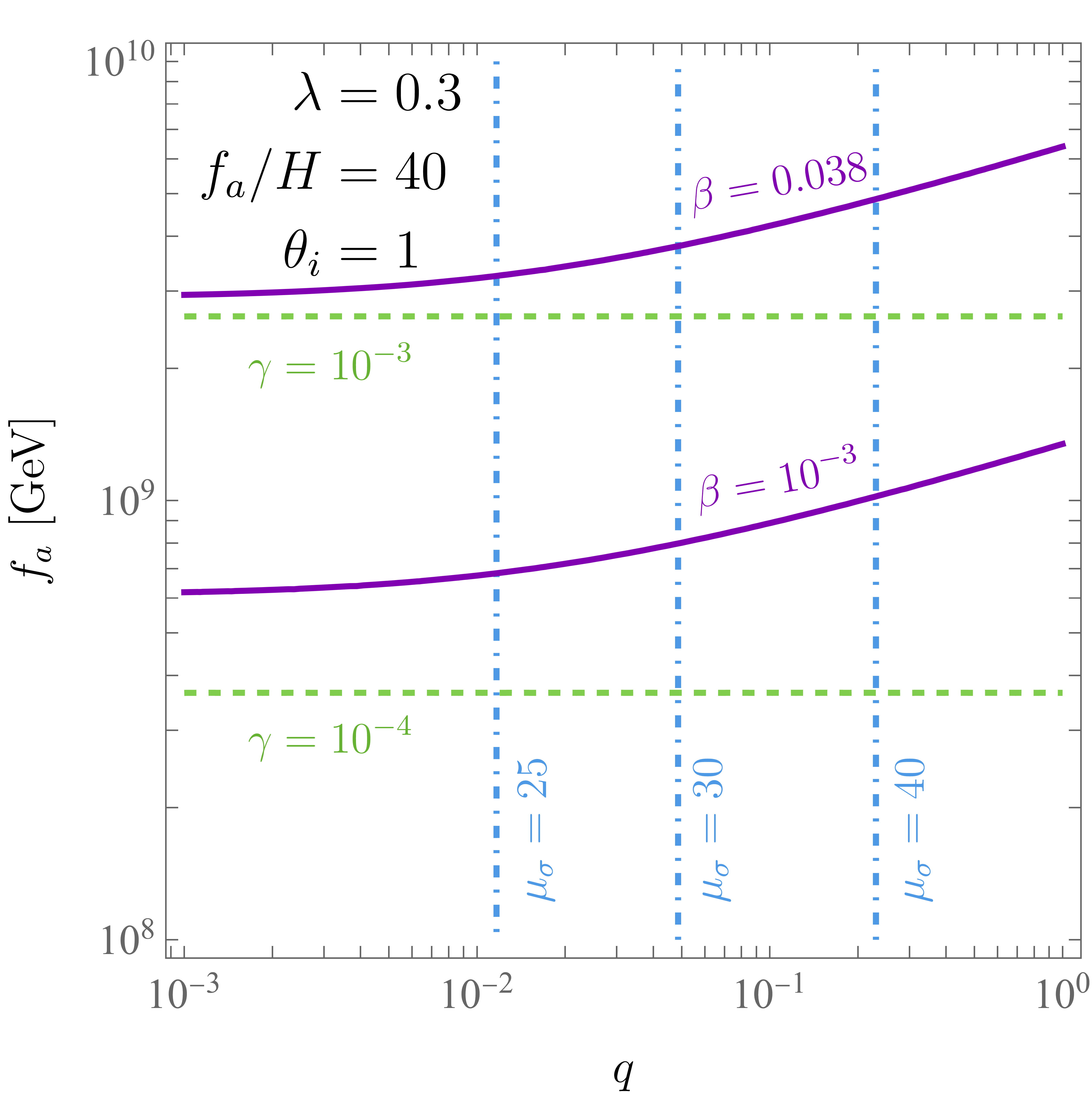}
    \includegraphics[width=7.5 cm]{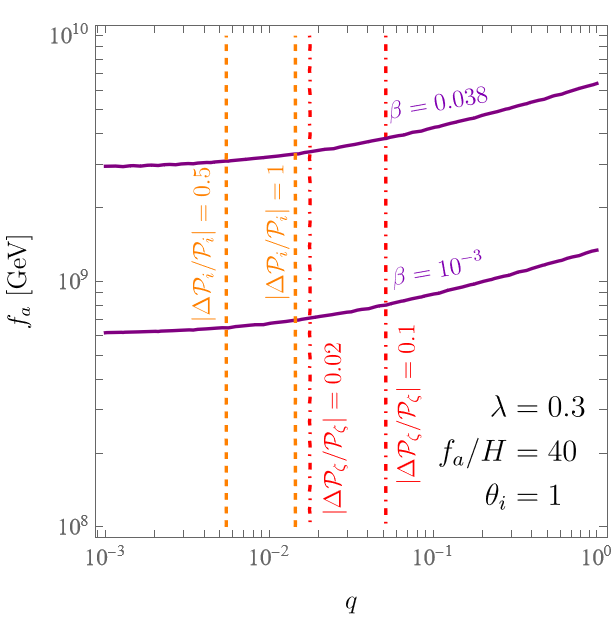} \\
      \includegraphics[width=7.5 cm]{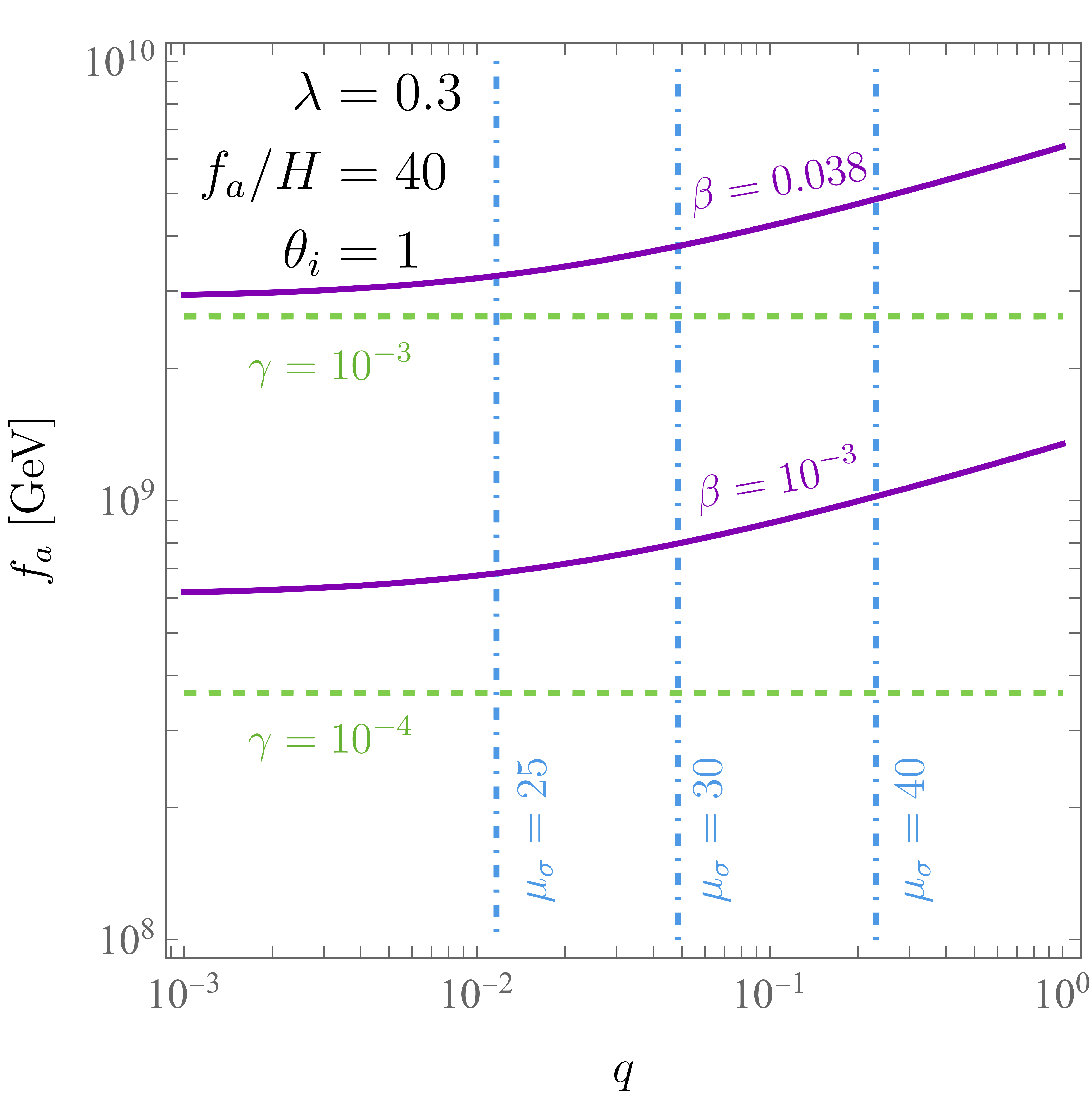}
    \includegraphics[width=7.5 cm]{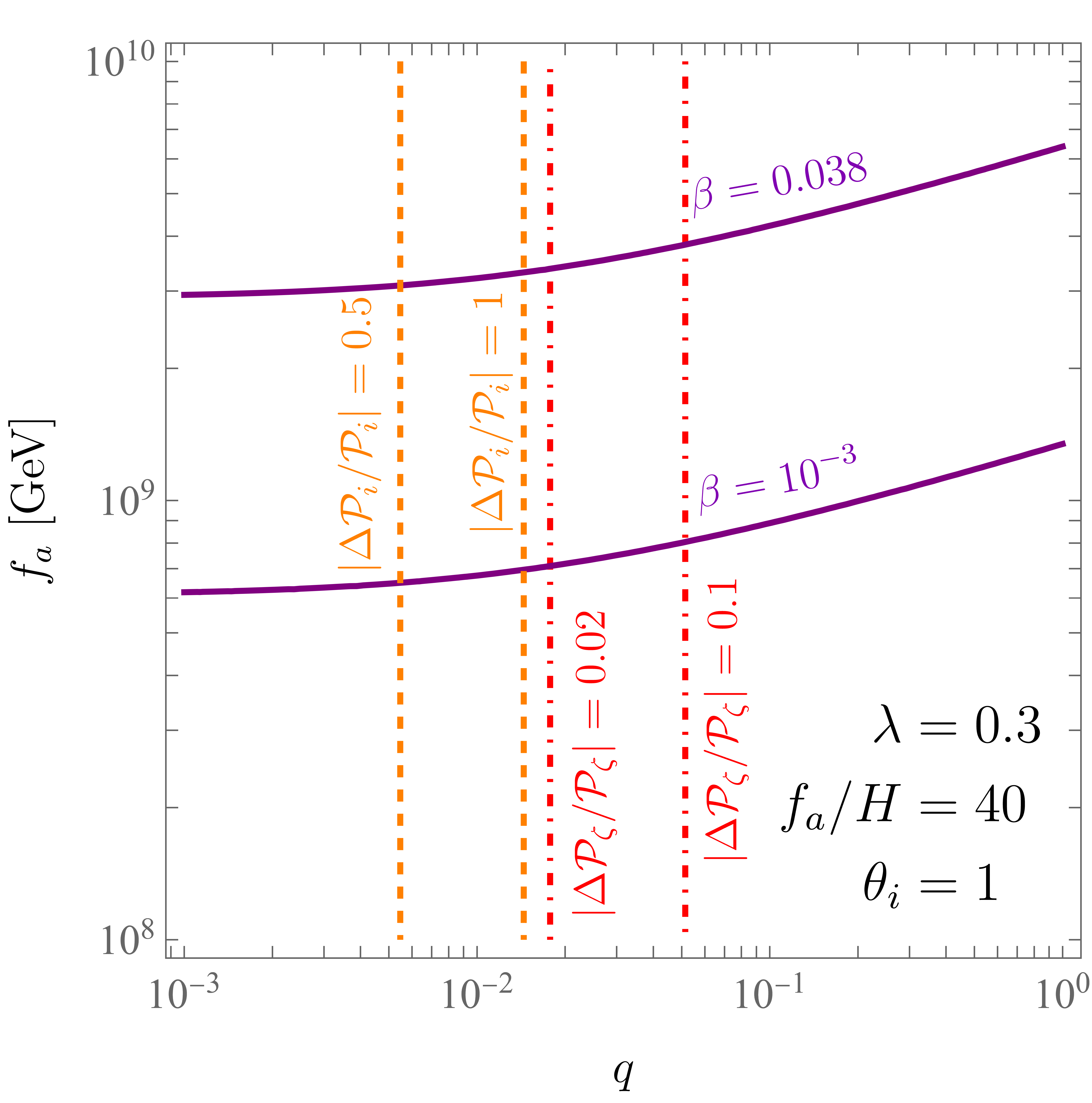}
    \caption{Left column: contours of the QCD axion DM fraction $\gamma$, isocurvature parameter $\beta$ defined in Eq.~\eqref{eq:isoconstraints}, and radial mode mass in Hubble unit, $\mu_\sigma$, in the $(q, f_a)$ plane. Right column: contours of $\left| \frac{\Delta P_\zeta}{P_\zeta}\right|_{\rm clock; amp}$ and $\left| \frac{\Delta P_i}{P_i}\right|_{\rm clock; amp}$, as well as $\beta$ in the same plane. Top row: $\lambda = 1$; bottom row: $\lambda = 0.3$. We fix $\theta_i =1$ and $f_a/H = 40$. For the two right panels, we also choose $b V_{\phi0}/\dot{\phi}_0^2 = 0.3$. 
    }
    \label{fig:model1par}
\end{figure}

Estimates in the more general parameter space are shown in Fig.~\ref{fig:model1par}. We fix the initial misalignment angle $\theta_i$ to be one and $\lambda$ to be an order one value, either 1 (top row) or 0.3 (bottom row), and present contours of physical quantities in the plane of the two most important input model parameters $f_a$ and $q$. We also fix $f_a/H = 40$, but the result won't change much when varying $f_a/H$ as long as it is ${\cal O}(10)$. For $f_a/H \gtrsim 100$, $\mu_\sigma = \sqrt{\lambda} f_I/H > \sqrt{\lambda} f_a/H$ also becomes $\gtrsim 100$ (for $\lambda \sim {\cal O}(1)$) and makes it more difficult for the clock signals to be observable in CMB due to the high frequency. Here we assume that the axion is the QCD axion, and the relic abundance is computed according to the last line of Eq.~\eqref{eq:isocurvatureanda}. Note that $\Omega_a \propto f_a^{\frac{7}{6}}$, a higher $f_a$ will give rise to a larger axion fraction $\gamma$. Combining the left and right panels, we see that, for example, when $bV_{\phi0}/\dot\phi_0^2=0.3$, $q \gtrsim 0.01$ and $f_a \lesssim$ a few $\times 10^9$ GeV,\footnote{$f_a \gtrsim 10^8$ GeV given the current constraints on the QCD axion. See the summary plot~\cite{AxionLimits}. }  $|\Delta P_\zeta/P_\zeta| \gtrsim 2\%$ with $\mu_\sigma$ of order a few 10's, which could be observable, while the current isocurvature constraint in Eq.~\eqref{eq:isoconstraints} is still satisfied. In this region, the axion DM fraction $\gamma \sim {\cal O}(10^{-3} - 10^{-4})$.

\begin{figure}[h!]
    \centering
    \includegraphics[width=10cm]{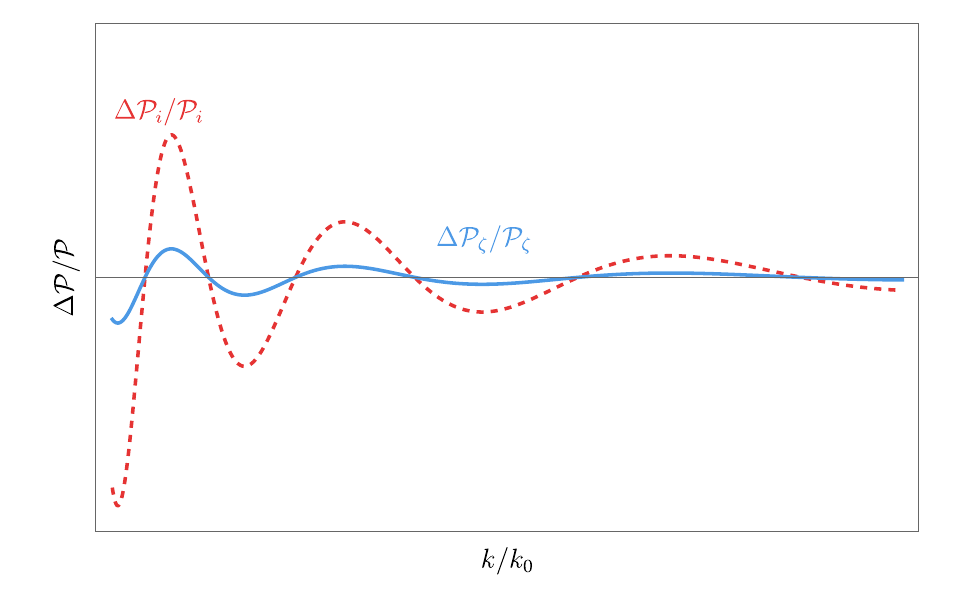}
    \caption{Illustration of the correlated clock signals in the power spectra in model 1: $\Delta P_i/P_i$ (red dashed) and $\Delta P_\zeta/P_\zeta$ (blue solid). Both axes are in linear scales. The decaying envelope is due to $(k/k_r)^{-3/2}$ as shown in Eq.~\eqref{eq:model1Pcclock} and Eq.~\eqref{eq:model1Piclock}.}
    \label{fig:correlatedclock}
\end{figure}

From the computations above, one can see that our model 1 could have a striking signal in the power spectra, namely the correlated clock signals, with the following characteristics: 
\begin{itemize}
\item Sizable corrections to both the curvature and isocurvature power spectra, both taking the form $\sin\left(\mu_\sigma \ln (k/k_r) + {\rm phase}\right)$ with a common mass parameter $\mu_\sigma$, a common reference scale $k_r$, and exactly the same phase.
\item The correction to the isocurvature two-point correlator is parametrically larger by a factor of $1/q$. 
\end{itemize}
This signal is illustrated in Fig.~\ref{fig:correlatedclock}. Such a signal, if measured, could be very informative:
\begin{itemize}
\item The correlated clock signals would point to an oscillating massive field coupled to both the inflaton and the dark matter field, and therefore serve as a more distinctive signature of the axion-inflaton Lagrangian in Eq.~\eqref{eqn:Lagrangian2} than other more generic predictions of CC physics associated with a massive field.

\item It could help us determine $\mu_\sigma$ (or equivalently, $m_{\sigma;\rm eff}$ in Hubble unit) and $q$, and learn about the energy scale related to the PQ field and its coupling to the inflaton. 

\item We could also narrow down the range of the inflation energy scale from these measurements in this particular model. There are degeneracies between $f_I$ and $H$ in two observables, namely, the amplitude of the axion isocurvature perturbation $\beta \propto [\gamma H/(f_I \theta_i)]^2 \propto \theta_i^2 H^2 f_I^{1/3} \left[1- (2q/\lambda)(\dot{\phi}_0^2/f_I^4) \right]^{7/6}$ (combining Eq.~\eqref{eq:isocurvatureanda},~\eqref{eq:isoconstraints} and \eqref{eq:fIfa}), and the mass of $\sigma$ in Hubble units $\mu_\sigma = \sqrt{\lambda} (f_I/H)$ from the frequency of the clock signals, assuming QCD axion dark matter with the misalignment mechanism. The measurement from either the axion isocurvature perturbations or the primordial standard clock/cosmological collider physics alone would not be able to determine $H$ over many orders of magnitude. However, these two degeneracy directions in the $f_I$-$H$ plane are quite orthogonal to each other. Combining these two observables (together with the determination of $q$ from the relative sizes of the curvature/isocurvature clock signals\footnote{The dependence of $\beta$ on $q$ is much weaker than the other two parameters, $H$ and $f_I$. Thus the parameter degeneracy is mostly between $H$ and $f_I$. Even without a precise determination of $q$, we could still infer both $H$ and $f_I$ in narrow ranges.}), we would have a very rare opportunity of being able to significantly narrow down the energy scale of the inflation model, better than either type of the experiments alone in terms of order of magnitude, especially for the low-scale (small-field) inflation models of which the tensor mode would be too small to observe.\footnote{There are uncertainties from undetermined parameters such as $\theta_i$ and $\lambda$, which we take to be of $\mathcal{O}$(1) in the statement above. Uncertainties when determining $H$ also come from other sources, such as those from the relic abundance computation of the QCD axion DM~\cite{Dine:2017swf}. Nonetheless, given our ignorance about the scale of the inflationary Hubble parameter (over many orders of magnitude) and the scarcity of possible observational windows into this important parameter, such a method could still lead to a major improvement of our knowledge.} 

\item Interestingly, if this clock signal would be observable in the curvature power spectrum in the near future, which means $\Delta P_\zeta/P_\zeta$ should be at least a few percent \cite{Braglia:2021sun, Braglia:2021rej}, the correlated oscillatory signal would be present at the leading order of the isocurvature power spectrum because $q\ll 1$. In certain parameter space, the oscillatory signal could even become the leading component of the isocurvature power spectrum over the scale-invariant one.\footnote{This also means that the perturbative theory for the isocurvature spectrum could break down, and finding a method to calculate the amplitude more precisely  becomes an interesting challenge.} When a scale-dependent oscillatory signal becomes important, the constraint on the isocurvature perturbation from the Planck data may need to be re-analyzed with oscillatory templates, because such a signal would not be picked up by a scale-invariant template.

\item Both clock signals are examples of classical primordial standard clock signals. The phases of their oscillations directly record the time dependence in the background inflationary $a(t)$, and therefore provide direct evidence for the inflationary scenario. We refer the readers to the following literature for detailed studies of this aspect \cite{Chen:2011zf, Chen:2011tu, Chen:2012ja, Battefeld:2013xka, Gao:2013ota, Noumi:2013cfa, Saito:2012pd, Saito:2013aqa, Chen:2014joa, Chen:2014cwa, Huang:2016quc, Domenech:2018bnf, Braglia:2021ckn, Braglia:2021sun, Braglia:2021rej, Bodas:2022zca}.

\end{itemize}

The most stringent constraints on primordial feature models so far are from the CMB data. Many feature models have been compared to Planck data and constrained using the curvature power spectrum and non-Gaussianities (see \cite{Slosar:2019gvt,Achucarro:2022qrl} for reviews). 
With the addition of many other experiments and different types of cosmological data, we expect rapid progress in this area in the near future.
Future improvements from observations of CMB E-mode polarization are forecasted in \cite{Braglia:2022ftm}. Those from galaxy surveys are studied in \cite{Huang:2012mr,Hazra:2012vs,Chen:2016vvw,Ballardini:2016hpi,Palma:2017wxu,LHuillier:2017lgm,Ballardini:2017qwq,Beutler:2019ojk,Ballardini:2019tuc,Debono:2020emh,Li:2021jvz,Ballardini:2022wzu,Chandra:2022utq}. 
More futuristically, one can also look forward to 21cm hydrogen line surveys~\cite{Chen:2016zuu,Xu:2016kwz} and stochastic gravitational wave background mapping \cite{Fumagalli:2020nvq,Braglia:2020taf,Bodas:2022zca}.
So, the type of features in the curvature power spectrum, such as the benchmark model we presented in this work, will be critically tested very soon. On the other hand, experimental prospects of primordial features in the isocurvature power spectrum are now an open question.

\subsection{Three-point correlators}
\label{ssec:model1threepoint}

It is well-known that primordial features boost NGs of inflation models \cite{Chen:2006xjb,Chen:2008wn, Flauger:2009ab, Flauger:2010ja, Chen:2010bka, Adshead:2011jq,Hazra:2012yn, Bartolo:2013exa,Fergusson:2014hya,Fergusson:2014tza}.
With the presence of the axion isocurvature during inflation, there could also be large NGs in three-point correlators involving isocurvature modes. We will compute both in this section. The relevant trilinear perturbation terms for leading-order bispectra include 
\begin{align}
{\cal L}_1^{(3)} \supset  &-\frac{1}{6}\frac{\partial^3 V_{\phi1}}{\partial \phi^3} (\delta \phi)^3 \nonumber \\
&+ \frac{1}{f_I}\left(1+ \frac{\sigma_{\rm bkg}}{f_I}\right) \delta \sigma \left((\delta \dot{a})^2 - \frac{1}{R^2} (\partial_i \delta a)^2\right)~.
\label{eq:model1trilinear}
\end{align}

\noindent \underline{\it Curvature bispectrum.} A sharp feature in time could not only generate a sinusoidal signal in the curvature power spectrum, but also in a curvature bispectrum (which we will denote as $ccc$) through the first trilinear coupling in Eq.~\eqref{eq:model1trilinear}:
\begin{equation}
  -\frac{1}{6}\frac{\partial^3 V_{\phi1}}{\partial \phi^3} (\delta \phi)^3 = -\frac{1}{6}  \frac{b V_{\phi 0} H^3 \tau \tau_s}{\dot{\phi}_0^3}\;\left(\tau_s\partial^2_\tau\delta(\tau-\tau_s) -\partial_\tau\delta(\tau-\tau_s)\right) (\delta \phi)^3~.
  \label{eq:Vphitripleprime}
\end{equation}

\begin{figure}[h!]
    \centering
    \includegraphics[width=8cm]{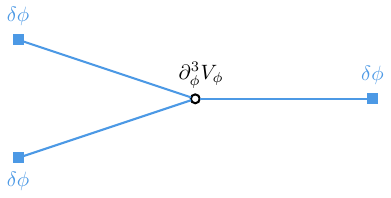}
    \caption{The curvature bispectrum induced by the sharp feature. The external points of inflaton are marked with blue squares. The solid blue line represents the propagator for $\phi$. The empty dot denotes the interaction in Eq.~\eqref{eq:Vphitripleprime}.  }
    \label{fig:cccclassical}
\end{figure}

 The corresponding diagram with an insertion of this interaction in Fig.~\ref{fig:cccclassical} could be computed as
\begin{align}
    &\langle \delta \phi^3 \rangle' \nonumber \\
    &\supset i \frac{-b V_{\phi0} H^3 \tau_s}{6\dot{\phi}_0^3} u_{k_1}u_{k_2}u_{k_3}(\tau_{\rm end})\int^0_{-\infty} \frac{\tau  d\tau}{(H \tau)^4}u^\ast_{k_1}u^\ast_{k_2}u^\ast_{k_3}(\tau) \left(\tau_s\partial^2_\tau\delta(\tau-\tau_s) -\partial_\tau\delta(\tau-\tau_s)\right) ~+~c.c.~,\nonumber\\
    & = \frac{b V_{\phi 0} H^5 }{4\dot{\phi}_0^3 } \left[g_1(k_1, k_2, k_3)\cos\bigg(\frac{k_1+k_2+k_3}{k_0}\bigg)+g_2(k_1, k_2, k_3)\sin\bigg(\frac{k_1+k_2+k_3}{k_0}\bigg) \right]~,
\end{align}
where $\langle \delta \phi^3 \rangle'$ is the normalized three-point inflaton correlator without the delta function: $\langle \delta \phi^3 \rangle \equiv \langle \delta \phi^3 \rangle' (2 \pi)^3 \delta^3 (\sum {\bf k}_i)$; $g_1, g_2$ are the model-dependent envelope functions. The $g_1, g_2$ functions from the simple step function potential we assume should be taken with a grain of salt since they may not be representative of more realistic models. Yet the amplitude still allows us to estimate the strength of NG measured by the dimensionless parameter of the $ccc$ correlator, $f_{\rm NL}^{ccc}$, 
\begin{align}
|f_{\rm NL}^{ccc}| \sim \frac{b V_{\phi 0}}{4\dot{\phi}_0^2 A_s^2} \left(\frac{H^2}{\dot{\phi}_0}\right)^4 = 4 \pi^4 \frac{b V_{\phi 0}}{\dot{\phi}_0^2 } ~,
\end{align}
where we use that the curvature perturbation $\zeta$ is related to $\delta \phi$ via $\zeta = - (H/\dot{\phi}_0) \delta \phi$ and the $ccc$ correlator is normalized by $A_s^2$. 
For $ b V_{\phi 0}/\dot{\phi_0^2} \sim {\cal O}(0.1)$, $|f_{\rm NL}^{ccc}|\sim {\cal O}(10-100)$.
Though it could be a large signal, this sharp-feature-induced $ccc$ NG does not tell us anything about the isocurvature mode, and we will not study it further. We also ignore high-order diagrams involving an exchange of the radial mode.

\begin{figure}[h!]
    \centering
    \includegraphics[width=9cm]{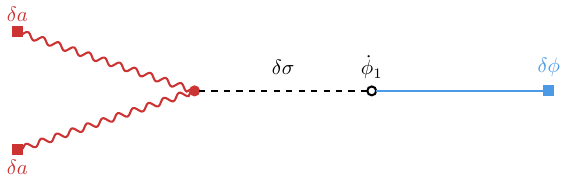}
    \caption{Diagram with an insertion of $\dot{\phi}_1$ (represented by an empty dot) to the $cii$ bispectrum in model 1. The filled red dot represents a constant vertex. }
    \label{fig:model1cii1}
\end{figure}

\begin{figure}[h!]
    \centering
    \includegraphics[width=\textwidth]{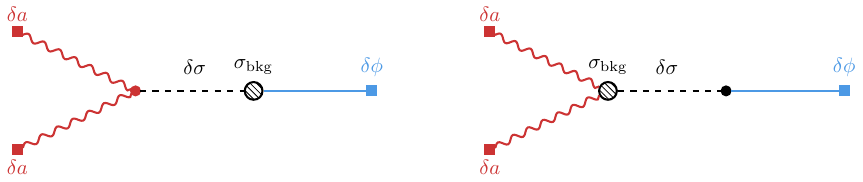}
    \caption{Diagrams with an insertion of $\sigma_{\rm bkg}$ (represented by a shaded blob) to the $cii$ bispectrum in model 1. The  filled dots (red or black) indicate that the vertices are constants.  }
    \label{fig:model1cii2}
\end{figure}

\noindent \underline{\it Mixed isocurvature-curvature bispectrum.} 
The quadratic couplings in Eq.~\eqref{eq:model1quadratic} and the trilinear ones in Eq.~\eqref{eq:model1trilinear} combined could contribute to curvature-isocurvature-isocurvature three-point correlators, which we will indicate as the ``$cii$" bispectrum. The leading types of diagrams are shown in Fig.~\ref{fig:model1cii1} and Fig.~\ref{fig:model1cii2}. Fig.~\ref{fig:model1cii1} has an insertion of $\dot{\phi}_1$, while the two diagrams in Fig.\ref{fig:model1cii2} have an insertion of the oscillating $\sigma_{\rm bkg}$ at either two-point or three-point vertex respectively. Note that in model 1, $\delta a$'s always come in pairs, so only bispectra with an even number of isocurvature modes exist. In other words, only $ccc$ or $cii$ exist.

In this section, we will compute the $cii$ three-point correlator with momenta $k_1, k_2, k_3$, where $k_{1,2}$ are taken to be associated with the axion modes and $k_3$ is assigned to the inflaton $\phi$. 
As we will see, the leading behavior of this correlator is a scale-dependent oscillatory function of the momentum $k_1+k_2+k_3$. Such a signal would be most easily detected in the equilateral limit where there are more triangle configurations.
So we will focus on the equilateral limit with $k_1=k_2=k_3 \equiv k$ in this section. Define the normalized three-point function (with a prime superscript) as:
\begin{equation}
    \langle \delta a(\textbf{k}_1) \delta a (\textbf{k}_2) \delta \phi(\textbf{k}_3)\rangle = (2\pi)^3 \delta^3(\textbf{k}_1+\textbf{k}_2+\textbf{k}_3)\langle\delta  a \delta a  \delta\phi\rangle'~.
\end{equation}
The contributions from Fig.~\ref{fig:model1cii1} and the left panel of Fig.~\ref{fig:model1cii2} include the following integral in common:
\begin{align}
\label{eq:threepoint2inj1}
   &u_{k_1} u_{k_2} u_{k_3}(\tau_{\rm end})   \int^0_{-\infty} \frac{d\tau_1}{(H \tau_1)^4} \partial_\mu u^\ast_{k_1} \partial^\mu u^\ast_{k_2} v_{k_3}(\tau_1)
    \int^{\tau_1}_{-\infty} \frac{d\tau_2}{(H \tau_2)^4}\bigg(\frac{\tau_2}{\tau_s}\bigg)^j \dot{u}^\ast_{k_3} v^\ast_{k_3}(\tau_2)\theta(\tau_2-\tau_s)~,
\end{align}
where $j=3$ or $\frac{3}{2}\pm i\mu_\sigma$ depending on whether $\dot{\phi}_1$ or $\sigma$ is injected. The massive mode function is given by $v_k(\tau)=-i e^{-\pi \mu_\sigma/2 + i\pi/4} \sqrt{\pi} H (-\tau)^{3/2} H_{i \mu_\sigma}^{(1)}(-k \tau)/2$ with $H_{i \mu_\sigma}^{(1)}$ the Hankel function of the first kind. This integral corresponds to having the feature injection at $\tau_2$ to excite the heavy $\sigma$ quantum before it decays to the axion quanta at $\tau_1$. 
The integration of the two-point vertex at $\tau_2$ reads:
\begin{align}
    &\int^{\tau_1}_{-\infty} \frac{d\tau_2}{(H \tau_2)^4}\bigg(\frac{\tau_2}{\tau_s}\bigg)^j \dot{u}^\ast_{k_3} v^\ast_{k_3}(\tau_2)\theta(\tau_2-\tau_s)\nonumber \\
    &=\int^{\tau_1}_{-\infty} d \tau_2\frac{\sqrt{\pi}(1+i) e^{\frac{\pi\mu_\sigma}{2}+i k_3 \tau_2} \sqrt{-k_3\tau_2}}{4 H\tau_2}\bigg(\frac{\tau_2}{\tau_s}\bigg)^j H^{(2)}_{i\mu_\sigma}(-k_3\tau_2) \theta(\tau_2-\tau_s)\nonumber \\
    &=\frac{\sqrt{\pi}(1+i) z_s^{-j}}{4 H}\int_{z_1}^{z_s} e^{\frac{\pi\mu_\sigma}{2}} e^{-i z_2} z_2^{j-\frac{1}{2}} H^{(2)}_{i\mu_\sigma}(z_2) dz_2~, 
     \label{eq:model1twopointint}
\end{align}
where $z_{1,2,s}=-k_3 \tau_{1,2,s} = -k \tau_{1,2,s}$, respectively. We also use the Hankel function of the second kind, $H_{i \mu}^{(2)}(z) = \left(H_{i\mu}^{(1)}(z)\right)^* e^{-\pi \mu}$. In the regime with $k/k_0\gtrsim \mu_\sigma \gg 1$ we are interested in, the early-time expansion of the Hankel function could be applied. The two-point integral could then be approximated as
\begin{align}
\label{eq:X2diagram1}
        &\int^{\tau_1}_{-\infty} \frac{d\tau_2}{(H \tau_2)^4}\bigg(\frac{\tau_2}{\tau_s}\bigg)^j \dot{u}^\ast_{k_3} v^\ast_{k_3}(\tau_2)\theta(\tau_2-\tau_s) \simeq \frac{-i (i z_s)^{-j}}{H 2^{1+q}}\bigg[\Gamma(j,2iz_s)-\Gamma(j,2iz_1)\bigg]\nonumber \\
       &\simeq \frac{e^{-2i z_1} z_s^{-j}  z_1^{j-1}}{4 H}\bigg[1-\frac{i(j-1)}{2z_1}\bigg]-\{z_1\to z_s\}~,
\end{align}
where $\Gamma(a,b)$ is the incomplete Gamma function.

The three-point vertex at $\tau_1$ has the following kinematic structure:
\begin{equation}
    u_{k_1} u_{k_2}(\tau_{\rm end})\partial_\mu u^\ast_{k_1} \partial^\mu u^\ast_{k_2}(\tau_1) = \frac{H^6\tau_1^2}{4 k_1^3 k_2^3} e^{i k_{12} \tau_1} [(\textbf{k}_1 \cdot \textbf{k}_2 )(k_1 k_2 \tau_1^2 + i k_{12} \tau_1 -1)-k_1^2 k_2^2 \tau_1^2]\equiv \frac{H^6\tau_1^2}{4 k_1^3 k_2^3} \mathcal{D}e^{i k_{12} \tau_1}~, 
\end{equation}
where $k_{12}\equiv k_1+k_2$ and the operator $\mathcal{D}$ is defined as 
\begin{align}
\mathcal{D} \equiv k_1^2 k_2^2 \partial_{k_{12}}^2 + (\textbf{k}_1 \cdot \textbf{k}_2)  \left(-1 +k_{12} \partial_{k_{12}} -k_1k_2 \partial_{k_{12}}^2\right)~,
\end{align}
with the partial derivative $\partial_{k_{12}}$ acting on $k_{12}$.
The integration involving the three-point vertex in the equilateral limit can be simplified as:
\begin{align}
\label{eq:threepointinf1}
        & u_{k_1} u_{k_2}(\tau_{\rm end})\int_{-\infty}^0\frac{ d\tau_1}{(H \tau_1)^4} \partial_\mu u^\ast_{k_1} \partial^\mu u^\ast_{k_2} v_{k_3}(\tau_1)\theta(\tau_1-\tau_s)=\int^0_{\tau_s} \frac{H^6 d\tau_1}{(H \tau_1)^4}\frac{\tau_1^2}{4 k_1^3 k_2^3} v_{k_3}(\tau_1) \mathcal{D}e^{i k_{12} \tau_1} \nonumber \\
        &=\frac{(-1)^{\frac{3}{4}}e^{-\pi \mu_{\sigma}/2}H^3\sqrt{\pi}}{8 k_1^3 k_2^3 k_3^{5/2}} \int_0^{z_s} dz_1 e^{-i k_{12}z_1/k_3}\left[(k_1^2k_2^2-\textbf{k}_1 \cdot \textbf{k}_2 k_1k_2)z_1^\frac{3}{2}+i \textbf{k}_1 \cdot \textbf{k}_2 k_{12}k_3 z_1^\frac{1}{2}+\textbf{k}_1 \cdot \textbf{k}_2 k_3^2/z_1^\frac{1}{2}\right]\nonumber \\
        &\times H^{(1)}_{i\mu_{\sigma}}(z_1), \nonumber \\
        &= \frac{(-1)^{\frac{1}{4}}e^{-\pi \mu_{\sigma}/2}H^3\sqrt{\pi}}{16 k^{9/2}} \int_0^{z_s} \frac{dz_1}{\sqrt{z_1}} e^{-2i z_1}(3i z_1^2 +2 z_1-i)  H^{(1)}_{i\mu_{\sigma}}(z_1),
\end{align}
where we set $\tau_{\rm end}=0$. In the last step, we apply the equilateral limit with $k_1=k_2=k_3=k$ and $\textbf{k}_1\cdot \textbf{k}_2=-k^2/2$. 

For convenience, we denote the integration of Eq.~\eqref{eq:threepoint2inj1} as:
\begin{equation}
    \int_0^{z_s} X_3(z_1) dz_1 \int_{z_1}^{z_s} X_2(z_2) d z_2~,
\end{equation}
where $X_{3}$ and $X_2$ are abbreviations of the three- and two-point vertexes, respectively. Combining Eq.~\eqref{eq:X2diagram1},~\eqref{eq:threepointinf1} and applying the early-time expansion in the integration of $X_3$, we have the following leading-order analytic approximation for the $j=3$ case ($\dot{\phi}_1$ injection in Fig.~\ref{fig:model1cii1}):
\begin{align}
       \int_0^{z_s} X_3(z_1) dz_1 \int_{z_1}^{z_s} X_2(z_2) d z_2\simeq  \frac{H^3}{32k^{6} } e^{-3i z_s} + {\cal O}(1/z_s)~.
\end{align}
One could see that this contribution is a sinusoidal function with frequency set by $1/\tau_s$.

 The computations are similar for the left panel of Fig.~\ref{fig:model1cii2}, aside from the different forms and amplitudes of the injected features. However, numerically we find that the NG from the $\sigma_{\rm bkg}$ injection is subdominant compared to that of the $\dot{\phi}_1$ injection. First, the amplitude of $|\sigma_{\rm bkg}|/f_I \sim q (\lambda \dot{\phi}_0^2/m_{\sigma, {\rm eff}}^4) (b V_{\phi0}/\dot{\phi}_0^2)$ is usually smaller than $\dot{\phi}_1/\dot{\phi}_0 = b V_{\phi0}/\dot{\phi}_0^2$ except for very light $\sigma$. Moreover, since the frequency of the injected feature, $\mu_\sigma$, is the same as the propagator's mass, the oscillating $\sigma_{\rm bkg}$ field cannot excite a propagating $\delta \sigma$ perturbation with the same mass and light fields simultaneously. 
 In contrast, the $\dot{\phi}_1$ insertion is approximately infinitely sharp by construction and is capable of exciting this massive field.\footnote{In realistic model building, the sharpness of the step is not infinite and could be at the same order of the mass of the heavy field. In such cases, depending on parameter values, the two diagrams could contribute comparably. }
 Finally, from the right panel of Fig.~\ref{fig:step}, one can see that $\sigma_{\rm bkg}$ is continuous, in contrast to $\dot{\phi}_1$. In this case, the feature of $\sigma_{\rm bkg}$ is not as sharp as that of $\dot{\phi}_1$. The final integrand is a highly oscillatory function with its both ends at $z=0$ and $z_s$ being zero, making the overall integration small.

Another contribution to the diagrams comes from the integral where the three-point vertex happens at $\tau_2$ prior to the two-point one at $\tau_1$, namely:
\begin{align}
   &u_{k_1} u_{k_2} u_{k_3}(\tau_{\rm end})   \int^0_{-\infty} \frac{d\tau_1}{(H \tau_1)^4} \bigg(\frac{\tau_1}{\tau_s}\bigg)^j \dot{u}^\ast_{k_3} v_{k_3}(\tau_1)\theta(\tau_1-\tau_s) 
    \int^{\tau_1}_{-\infty} \frac{d\tau_2}{(H \tau_2)^4}\partial_\mu u^\ast_{k_1} \partial^\mu u^\ast_{k_2} v^\ast_{k_3}(\tau_2)~,
\end{align}
or in the shorthand notation $\int_0^{\infty} \tilde{X}_2(z_1) d z_1 \int_{z_1}^{\infty}  \tilde{X}_3(z_2) dz_2$, where $\tilde{X}_{2,3}$ stands for the two- and three-point vertexes. Numerically, we find the size of the integral above is subdominant or, at most, comparable to that of Eq.~\eqref{eq:threepoint2inj1}.
Similar to what is explained in \cite{Chen:2022vzh} (for the diagrams Fig.~2b and 2c in \cite{Chen:2022vzh}), the suppression is because this integral corresponds to the situation in which the massive field is created before the time of sharp feature.

We now estimate the size of contributions from Fig.~\ref{fig:model1cii1} to $cii$ bispectrum, up to a numerical factor, as
\begin{align}
   \left| \langle \delta a \delta a\delta \phi\rangle'\right| &\supset \frac{2 c \dot{\phi}_0}{\Lambda^2} \frac{H^3}{32 k^6} \frac{\dot{\phi}_1}{\dot{\phi}_0} = \frac{2 c \dot{\phi}_0}{\Lambda^2}\frac{H^3}{32 k^6} \frac{b V_{\phi0}}{\dot{\phi}_0^2} ~.
\end{align}
The three-point function above can be translated into the dimensionless form of NG parameter, $f_{\rm NL}^{cii}$, via the following relation:
\begin{align}
\label{eq:ciianalytical1}
    &\frac{k_1^2 k_2^2 k_3^2}{A_s A_i} \left|\langle  S_d(k_1)  S_d(k_2)  \zeta(k_3)\rangle'\right| \equiv \left|f_{\rm NL}^{cii}\right|~, 
 \nonumber\\
    &\left|\langle  S_d(k_1)  S_d(k_2)  \zeta(k_3) \rangle'\right| \simeq  \frac{4 \gamma^2 H}{f_I^2 \theta_i^2 \dot{\phi}_0} \left|\langle \delta a \delta a \delta \phi \rangle' \right|\simeq \beta \frac{8\pi^3 A_s^{3/2}}{H^3} \left|\langle \delta a \delta a \delta \phi \rangle'\right|~,
\end{align}
where we use $S_d = \gamma S_a =2 \gamma \delta a/(f_I \theta_i)$, $\zeta = - (H/\dot{\phi}_0) \delta \phi$ and the definition of $\beta$ in Eq.~\eqref{eq:isoconstraints}. Normalizing to the standard curvature NG parameter to take into account that the isocurvature power is suppressed, the contribution from Fig.~\ref{fig:model1cii1} to the $cii$ NG in the equilateral limit is estimated to be:
\begin{align}
   |f_{\rm NL}^{cii}| \frac{A_i}{A_s} & \supset \frac{A_i}{A_s} \frac{ 8\beta\pi^3 A_s^{3/2}}{H^3}\frac{k^6}{A_s A_i}\frac{ b V_{\phi0} H^3 c }{16 \Lambda^2 k^6 \dot{\phi}_0}=\frac{b V_{\phi_0}}{\dot{\phi}_0^2}\frac{\dot{\phi}_0 \pi^3 c\beta}{2 \sqrt{A_s}\Lambda^2}~, \nonumber \\
   &\simeq 40\left(\frac{b V_{\phi_0}}{0.3\dot{\phi}_0^2}\right)\left(\frac{\dot{\phi}_0}{(60H)^2}\right) \left(\frac{q}{0.02}\right) \left(\frac{\beta}{0.01}\right)  \left(\frac{40H}{f_I}\right)^2~.
\end{align}

The right panel of Fig.~\ref{fig:model1cii2} also contributes to $cii$ through the integration, in the equilateral limit:
\begin{align}
\label{eq:threepoint3inj1}
    &u_{k_1} u_{k_2} u_{k_3}(\tau_{\rm end})   \int^0_{-\infty} \frac{d\tau_1}{(H \tau_1)^4}\dot{u}^\ast_{k_3} v_{k_3}(\tau_1)
    \int^{\tau_1}_{-\infty} \frac{d\tau_2}{(H \tau_2)^4} \bigg(\frac{\tau_2}{\tau_s}\bigg)^j \partial_\mu u^\ast_{k_1} \partial^\mu u^\ast_{k_2} v^\ast_{k_3}(\tau_2)\theta(\tau_2-\tau_s)\nonumber \\
    &=\frac{-H^3 \pi}{64 k_3^6 } \int^{z_s}_0  d z_1 \frac{e^{-i z_1}}{\sqrt{z_1}}H^{(1)}_{i\mu_\sigma}(z_1)\int_{z_1}^{z_s} e^{-2i z_2}d z_2 \bigg(\frac{z_2}{z_s}\bigg)^{j} (3 z_2^\frac{3}{2}-2i z_2^\frac{1}{2}-z_2^{-\frac{1}{2}}) H^{(2)}_{i\mu_\sigma}(z_2) \nonumber \\
    &=\frac{-H^3 \pi}{64 k_3^6 } \int_{0}^{z_s} e^{-2i  z_2}d z_2 \bigg(\frac{z_2}{z_s}\bigg)^{j} (3 z_2^\frac{3}{2}-2i z_2^\frac{1}{2}-z_2^{-\frac{1}{2}}) H^{(2)}_{i\mu_\sigma}(z_2)  \int^{z_2}_0  d z_1 \frac{e^{-i z_1}}{\sqrt{z_1}}H^{(1)}_{i\mu_\sigma}(z_1)~,
\end{align}
where $\sigma_{\rm bkg} = \sigma_1$ with $j=\frac{3}{2}\pm i\mu_\sigma$ is injected through the three-point vertex at time $\tau_2$ prior to $\tau_1$. With the early-time expansion, the integration of the two-point vertex over $z_1$ reads:
\begin{align}
    e^{-\frac{\pi\mu_\sigma}{2}} \int^{z_2}_0  d z_1 \frac{e^{-i z_1}}{\sqrt{z_1}}H^{(1)}_{i\mu_\sigma}(z_1) \simeq  \frac{(1-i) \log z_2}{\sqrt{\pi}}~.
\end{align}
However, when combined with the integration of the three-point vertex, the logarithmic behavior above makes it difficult to obtain an analytical approximation. Moreover, the positive- and negative-frequency parts of the $\sigma_{\rm bkg}$ oscillation largely cancels with each other, which is also challenging to be approximated analytically. To have a rough estimation in the early-time regime, it is convenient to replace $\log z_2$ by a constant such as 1. With such a simplification, the integration over the three-point vertex becomes a product of polynomials and sinusoidal oscillations. By comparing with the numerical integration of Eq.~\ref{eq:threepoint3inj1}, we find the closed-form approximation of the $cii$ bispectrum to be
\begin{equation}
     \langle \delta a \delta a\delta \phi\rangle' \simeq -\frac{2 c \dot{\phi}_0}{\Lambda^2} \frac{|\sigma_1|}{f_I}  \frac{H^3 e^{-3iz_s}}{72 k^6} \log \left(\frac{2k}{k_r}\right)~.
\end{equation}
  Notice that the approximation above tends to overestimate the three-point function when $\mu_\sigma$ is large. When $z_s \gg \mu_\sigma \gg 1$, this contribution to the size of $f_{\rm NL}^{cii}$ could be estimated as
\begin{align}
\label{eq:ciianalytical2}
    |f_{\rm NL}^{cii}| \frac{A_i}{A_s} & \supset \frac{A_i}{A_s} \frac{ 8\beta\pi^3 A_s^{3/2}}{H^3}\frac{k^6}{A_s A_i} \frac{|\sigma_1|}{f_I} \frac{2c \dot{\phi}_0}{\Lambda^2} \frac{H^3}{72k^6 } \simeq \frac{b V_{\phi_0}}{\dot{\phi}_0^2} \frac{4 \pi^3 c^2 \dot{\phi}_0^3 \beta}{9 \sqrt{A_s} H^2 \Lambda^4 \mu_\sigma^2}\nonumber \\
    & \simeq \frac{4}{\lambda} \left(\frac{q}{0.02}\right)^2 \left(\frac{b V_{\phi0}}{0.3 \dot{\phi}_0^2} \right) \left(\frac{\dot{\phi}_0}{(60H)^2}\right)^3 \left(\frac{40 H}{f_I}\right)^{6} \left(\frac{\beta}{0.01}\right).
\end{align}

\begin{figure}[hbt!]
    \centering
    \includegraphics[height=8 cm]{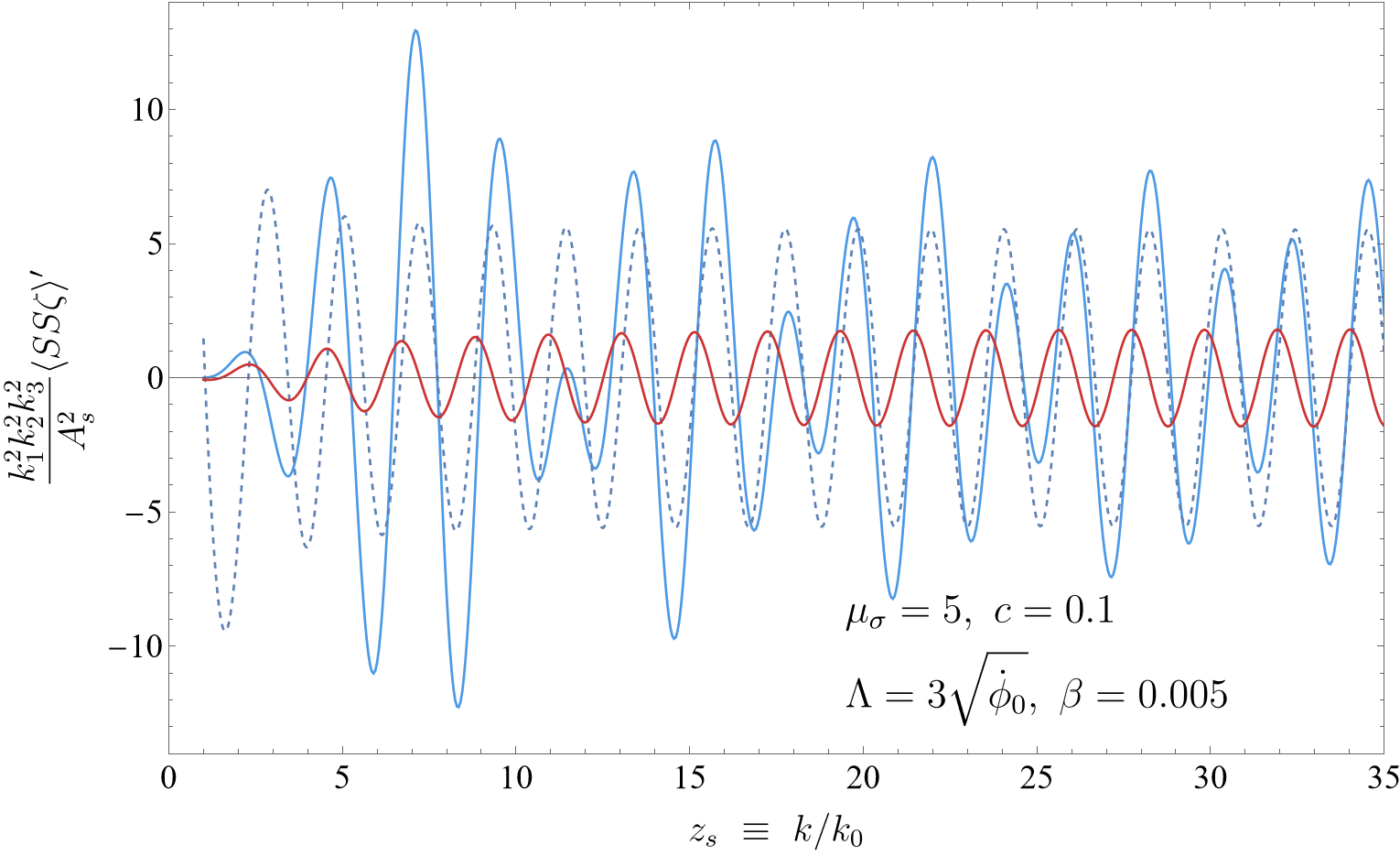}
    \includegraphics[height=8 cm]{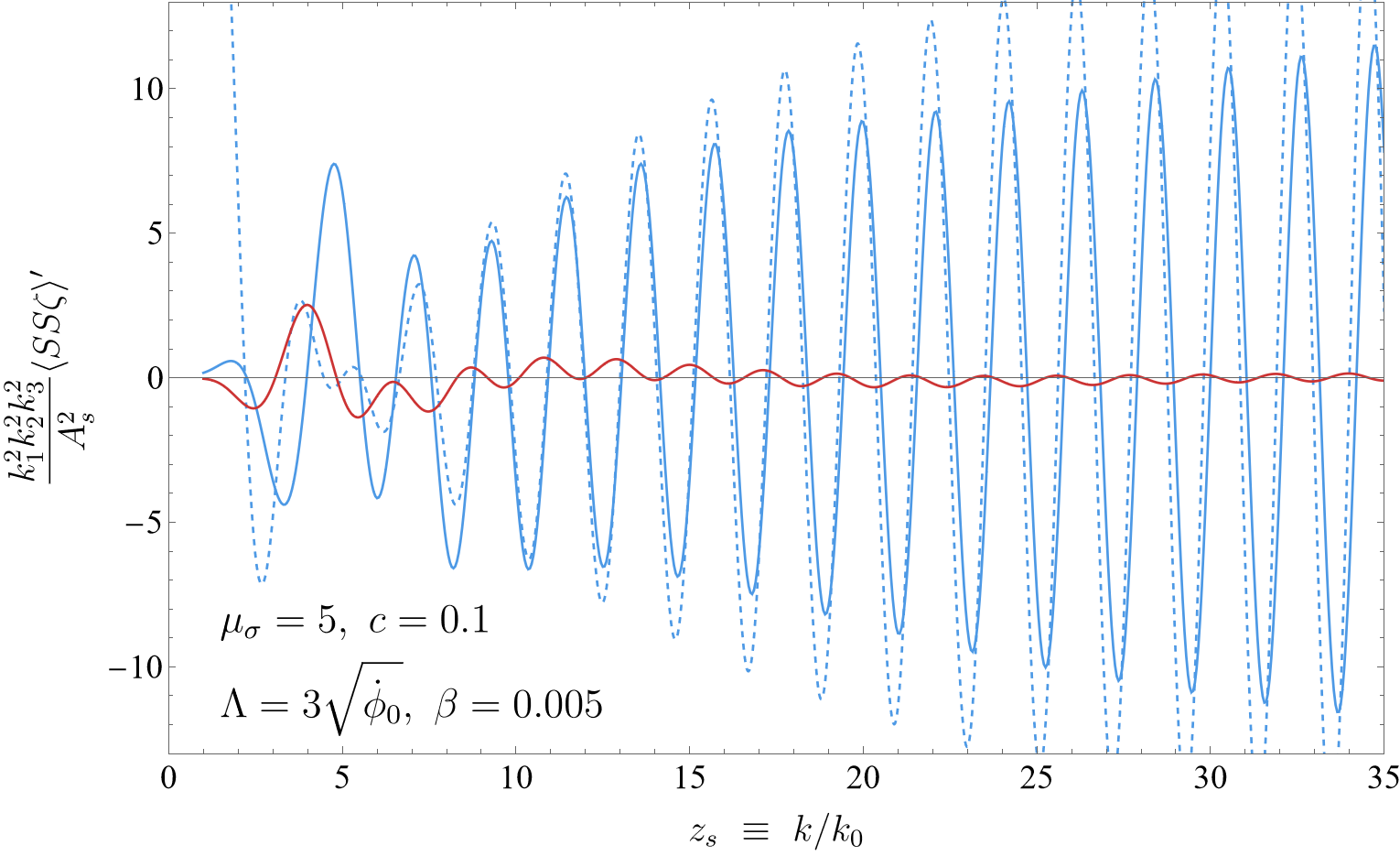}
    \caption{Sizes of the $cii$ NG for two different types of diagrams in the equilateral limit. We set $\mu_\sigma=5$, $c=0.1$, $\Lambda=3\sqrt{\dot{\phi}_0}$, and $\beta=5\times 10^{-3}$. In both panels, blue solid curves stand for diagrams where the feature insertions happen first, while red solid curves are the contributions of alternative diagrams where the features are injected at a later time. The blue dashed curves show the analytical approximation of the corresponding solid curves. Top: the diagram in Fig.~\ref{fig:model1cii1} with $\dot{\phi}_1$ injection. Bottom: the diagram with the $\sigma_{\rm bkg}$ injection at the three-point vertex, as shown in the right panel of Fig.~\ref{fig:model1cii2}. }
    \label{fig:ciiSbenchmark1}
\end{figure}

In Fig.~\ref{fig:ciiSbenchmark1}, we show the numerical results for a benchmark model with a relatively small $\mu_\sigma=5$ in the equilateral limit. For both types of diagrams in Fig.~\ref{fig:model1cii1} and Fig.~\ref{fig:model1cii2} (right panel), the $cii$ signals are sizable. The analytical approximations in Eq.~\eqref{eq:ciianalytical1} and~\eqref{eq:ciianalytical2} using the early-time expansion are also plotted as dashed curves, agreeing decently well with the numerical results in both the amplitude and the phase. For more realistic benchmarks with higher $\mu_\sigma$ (which are difficult to be evaluated numerically due to the slow convergence), we expect that the diagrams with $\dot{\phi}_1$ injection are not significantly affected as their amplitudes are independent of $\mu_\sigma$ in the early-time region. In contrast, the importance of the diagrams with $\sigma_{\rm bkg}$ injection will decrease with a larger $\mu_\sigma$ as their amplitudes $\propto |\sigma_1|$ are suppressed by $\mu_\sigma^{-2}$. In all cases, the phase of the $cii$ signal follows $e^{-3i k/k_0} = e^{i(k_1+k_2+k_3)\tau_s}$ up to a constant shift.

In the other interesting limit, the squeezed limit with $k_1 \sim k_2 \gg k_3$, the NG signals due to the sharp feature could also be sizable, which we check by numerical computations. The properties of the contributions from each type of diagram, such as the relative importance, are similar to those in the equilateral limit. Numerical integration also shows that the $k_3/k_0$ dependences of signals are similar to their equilateral-limit counterparts. However, it is more difficult to find analytical approximations for both the amplitude and the phase of the signal, which we will not detail here.  

The sharp feature signals in the curvature bispectrum and mixed bispectrum studied in this subsection, as well as the sharp feature signal \eqref{eq:Vpp} in the power spectrum, are all due to the same sharp feature in the model, and so they all run sinusoidally (with $2k$ versus $k_1+k_2+k_3$ between power spectrum and bispectra) with the same starting scale, parameterized by $k_0$. Besides these leading behaviors, the more detailed envelope behavior and subleading phase shifts are more complicated and depend on the nature of the sharp feature, but correlations between different spectra may still deserve further studies.
In contrast to the clock signals, as mentioned, only from these sharp feature signals we will not be able to learn any specific information about the massive radial mode; thus these signals are less distinctive as signatures of the PQ field.
These comments also apply to the sharp feature signals in the next model.

Regarding the observational prospect of these scale-dependent oscillatory bispectra, we note that, although we have shown that the sizes of these bispectra are greatly enhanced relative to their attractor values due to the features, it remains an open question of how large these bispectra have to be to become observable in future experiments. There are some constraints on such curvature bispectra from the Planck data \cite{Fergusson:2014hya, Fergusson:2014tza, Planck:2018jri}.\footnote{In general, there exists some simple but not-yet-constrained shapes of curvature NGs~\cite{Freytsis:2022aho}.} In addition, the Planck data also sets some weak constraints on mixed curvature/isocurvature bispectra assuming local shapes, which do not apply to the shapes in our model~\cite{Planck:2019kim}. The future observational prospect is less studied. On the other hand, both the constraint and the future prospect of the oscillatory mixed bispectra are open questions. We leave these questions to future works.

\section{Model 2: $\partial \phi \partial |\chi|^2$}
\label{sec:model2}

In model 2, we consider the following Lagrangian, which introduces a kinetic mixing between the inflaton and the radial mode, after the PQ symmetry breaking: 
\begin{align}
 {\cal L}_2 &= -\frac{(\partial_\mu \phi)^2}{2} - |\partial_\mu \chi|^2 - V_\phi(\phi)- V_\chi(\chi) - \frac{\tilde{c} }{\Lambda}\partial_\mu \phi \,\partial^\mu |\chi|^2~\nonumber \\
 &= -\frac{(\partial_\mu \phi)^2}{2} - \frac{(\partial_\mu \sigma)^2}{2}- \frac{1}{2}\left(1+\frac{\sigma}{f_I}\right)^2(\partial_\mu a)^2 - V_\phi(\phi)- V_\sigma(\sigma) - \frac{\tilde{c} }{\Lambda}\partial_\mu \phi \, \partial^\mu \sigma \left(f_I + \sigma\right) ~\nonumber \\
 &=-\frac{(\partial_\mu \tilde{\phi})^2}{2} - \frac{(\partial_\mu \tilde{\sigma})^2}{2}- \frac{(\partial_\mu a)^2}{2} - V_\phi\left(\tilde{\phi}-\frac{x}{\sqrt{1-x^2}}\tilde{\sigma}\right)- V_\sigma\left(\frac{\tilde{\sigma}}{\sqrt{1-x^2}}\right) \nonumber\\ 
 & -\frac{\tilde{\sigma}(\partial_\mu a)^2}{\sqrt{1-x^2}f_I} -\frac{\tilde{\sigma}^2(\partial_\mu a)^2}{2(1-x^2)f_I^2} - \frac{x\tilde{\sigma}}{(1-x^2)f_I} \partial_\mu \tilde{\sigma}\, \partial^\mu\left(\tilde{\phi} - \frac{x}{\sqrt{1-x^2}}\tilde{\sigma}\right)~,
 \label{eq:fullmodel1}
\end{align}
where in the second line, we use the parametrization in Eq.~\eqref{eq:chiparametrization} and Eq.~\eqref{eq:Vrho}. 
In the third and fourth lines, we employ the following field redefinition to remove the kinetic mixing $\frac{\tilde{c} f_I}{\Lambda}\partial_\mu \phi \, \partial^{\mu} \sigma$:
\begin{equation}
    \tilde{\phi}\equiv \phi + x\sigma~,~ \tilde{\sigma} \equiv \sqrt{1-x^2} \sigma~,
\end{equation}
where the dimensionless parameter $x$ is defined as
\begin{equation}
    ~x\equiv \frac{\tilde{c} f_I}{\Lambda} \ll 1~.
\end{equation}
For the redefined canonically normalized fields, we could use the standard Bunch-Davies solutions for the mode functions.

In principle, the dimension-five operator added here could appear together with the dimension-six operator considered in the previous section. Since they lead to different signals and may come from different UV origins, we study them one by one for clarity and simplicity. As we will see, the signal magnitudes are not determined purely by the operator dimensions.

In the absence of a primordial feature, the homogeneous background solution of the inflaton is the same as in Eq.~\eqref{eq:slowroll} with $\phi_0$ replaced by $\tilde{\phi}_0$. Around the PQ field's VEV $f_I$ during inflation, the background radial mode satisfies
\begin{equation}
\tilde{\sigma}_0 = 0~,
\label{eq:rho0}
\end{equation}
which requires 
\begin{equation}
  \left.  \frac{\partial\left[V_\phi \left(\tilde{\phi}-\frac{x}{\sqrt{1-x^2}}\tilde{\sigma}\right)+V_\sigma \left(\frac{\tilde{\sigma}}{\sqrt{1-x^2}}\right)\right]}{\partial \tilde{\sigma}}\right|_{\tilde{\phi}=\tilde{\phi}_0, \tilde{\sigma}=0}=0~.
\end{equation}
Using Eq.~\eqref{eq:Vrho} and Eq.~\eqref{eq:slowroll}, one obtains a relation determining $f_I$ as well as the effective mass squared of $\tilde{\sigma}$:
\begin{equation}
    \lambda (f_I^2-f_a^2)f_I -6 x H \dot{\tilde{\phi}}_0 =0~,~\rightarrow~f_I \approx f_a + 3 x \frac{H \dot{\tilde{\phi}}_0}{\lambda f_a^2} + {\cal O}(x^2),
\end{equation}
and 
\begin{equation}
    m_{\tilde{\sigma},\rm{eff}}^2 \approx \left.\frac{\partial^2 V_\sigma (\tilde{\sigma})}{\partial \tilde{\sigma}^2}\right|_{\tilde{\sigma}=0} = \frac{\lambda}{2(1-x^2)}(3 f_I^2-f_a^2)\approx \lambda f_a^2 + 9x \frac{H \dot{\tilde{\phi}}_0}{f_a} + {\cal O}(x^2) ~,
\end{equation}
where we ignore the suppressed contribution to $m_{\tilde{\sigma},\rm{eff}}^2$ from $\partial^2 V_\phi/ \partial \tilde{\sigma}^2 \sim x^2 (\partial^2 V_\phi/ \partial \tilde{\phi}^2)$. The axion field adopts a trivial background solution $a_0(t) = 0$.

Now as in model 1, we add a step feature, $V_{\phi1}$ as in Eq.~\eqref{eq:Vphi11}, to the inflationary potential that sources the attractor solution, $V_{\phi0}$. The homogeneous background fields become
\begin{equation}
\tilde{\phi}_{\rm bkg}(t) = \tilde{\phi}_0(t)+\tilde{\phi}_1(t)~,~~
\tilde{\sigma}_{\rm bkg} (t) = \tilde{\sigma}_0(t)+\tilde{\sigma}_1(t) = \tilde{\sigma}_1(t)~,
\end{equation}
where $\tilde{\phi}_0(t)$ and $\tilde{\sigma}_0(t)$ are the featureless solutions as discussed above while $\tilde{\phi}_1(t), \tilde{\sigma}_1(t)$ are the induced features in the inflaton and radial mode trajectories. The axion field is not directly coupled to the inflaton and its background evolution remains trivial with $a_{\rm bkg}(t)=a_0(t)$ as a constant.
The linearized EOMs of $\tilde{\phi}_1(t)$ and $\tilde{\sigma}_1(t)$ become:
\begin{equation}
    \ddot{\tilde{\phi}}_1 +3 H\dot{\tilde{\phi}}_1 + \left.\frac{\partial{V_{\phi 1}}(\tilde{\phi})}{\partial\tilde{\phi}}\right|_{\tilde{\phi}=\tilde{\phi}_0} \simeq 0~, 
\end{equation}
and 
\begin{align}
 &\ddot{\tilde{\sigma}}_1 +3 H\dot{\tilde{\sigma}}_1 -\left.\frac{x}{\sqrt{1-x^2}} \frac{\partial{V_{\phi 1}}(\tilde{\phi})}{\partial\tilde{\phi}}\right|_{\tilde{\phi}=\tilde{\phi}_0} + m_{\tilde{\sigma},\rm{eff}}^2 \,\tilde{\sigma}_1 \simeq 0~. 
\end{align}
Plugging in Eq.~\eqref{eq:Vphi11}, the solutions of $\tilde{\phi}_1, \tilde{\sigma}_1$ read:
\begin{align}
\tilde{\phi}_1(t) &= \frac{b  V_{\phi 0}}{3 H \dot{\tilde{\phi}}_0} \left[1-e^{-3H(t-t_s)}\right]\theta(t-t_s)~, \label{eq:dim6featuresphi1} \\
    \tilde{\sigma}_1(t) &\simeq -\frac{x \, b V_{\phi 0}}{{\dot{\tilde{\phi}}_0}\, m_{\tilde{\sigma},{\rm eff}}}e^{-\frac{3}{2}H(t-t_s)} \sin \left[\mu_{\tilde{\sigma}} H (t-t_s)\right] \theta(t-t_s)~,~~\mu_{\tilde{\sigma}} \equiv \sqrt{\frac{m^2_{\tilde{\sigma},{\rm eff}}}{H^2} -\frac{9}{4}}~.
    \label{eq:dim6featuresrho1} 
\end{align}
Compared with the solutions in model 1 shown in Eqs.~\eqref{eq:dim5featuresphi1} and~\eqref{eq:dim5featuresrho1}, the induced inflaton feature remains the same while the induced oscillating radial mode is suppressed by one less power of its mass.

\subsection{Corrections to the power spectra}
\label{ssec:model2twopoint}
Let's first consider the correction to the power spectra. At the quadratic level, the interactions between the perturbations, $\delta \tilde{\phi}=\tilde{\phi}-\tilde{\phi}_{\rm bkg}$, $\delta \tilde{\sigma}=\tilde{\sigma}-\tilde{\sigma}_{\rm bkg}$ and $\delta \tilde{a}=\tilde{a}-\tilde{a}_{\rm bkg}=\tilde{a}$, read
\begin{align}
    \mathcal{L}_{2}^{(2)} &\supset 
    - \frac{1}{2}\frac{\partial^2 V_\phi}{\partial \tilde{\phi}^2} \left(\delta \tilde{\phi}-\frac{x}{\sqrt{1-x^2}}\delta\tilde{\sigma}\right)^2 -  \frac{1}{2}\frac{\partial^2 V_\sigma}{\partial \tilde{\sigma}^2}\delta\tilde{\sigma}^2 \nonumber \\
    &+ \left(\frac{\tilde{\sigma}_{\rm bkg}}{\sqrt{1-x^2}f_I}+\frac{\tilde{\sigma}_{\rm bkg}^2}{2(1-x^2)f_I^2}\right)\left(\delta\dot{a}^2-\frac{1}{R^2}(\partial_i \delta a)^2\right)\nonumber \\
    &+\frac{x}{(1-x^2) f_I}\bigg(\dot{\tilde{\phi}}_{\rm bkg} \delta \dot{\tilde{\sigma}}\delta\tilde{\sigma}+\dot{\tilde{\sigma}}_{\rm bkg} \delta \tilde{\sigma}\delta \dot{\tilde{\phi}}+\tilde{\sigma}_{\rm bkg} \delta \dot{\tilde{\sigma}}\delta \dot{\tilde{\phi}} - \frac{\sigma_{\rm bkg}}{R^2} (\partial_i \delta \tilde{\phi}) \, (\partial_i \delta\tilde{\sigma}) \bigg)~. 
  \label{eq:model2quadratic}
\end{align}

\begin{figure}[h!]
    \centering
    \includegraphics[width=10cm]{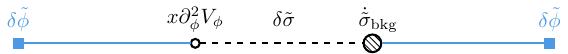}
    \caption{An example of a higher-order diagram contributing to the curvature power spectrum with the exchange of the heavy radial mode. The external points of inflaton are marked with blue squares. The solid blue and dashed black lines represent the propagator for $\tilde{\phi}$ and $\tilde{\sigma}$, respectively. The empty dot denotes the interaction proportional to $x \partial^2_\phi V_{\phi}$ in the first line of Eq.~\eqref{eq:model2quadratic}, while the shaded blob represents interactions with an insertion of $\dot{\tilde{\sigma}}_{\rm bkg}$ in the third line of the same equation.  }
    \label{fig:model2cc}
\end{figure}

The first term in Eq.~\eqref{eq:model2quadratic}, $-\frac{1}{2}\frac{\partial^2 V_\phi}{\partial \tilde{\phi}^2} (\delta \tilde{\phi})^2$, due to the step potential $V_{\phi1}$ is also present in model 1. It will lead to the same sharp feature signal in the curvature spectrum, computed in Eq.~\eqref{eq:correctioncur2pt1}. We will not repeat the computation here.
On the other hand, in contrast to model 1, there are no quadratic terms of $(\delta \tilde{\phi})^2$ from the dimension-5 operator we add in model 2 since it is only linear in $\tilde{\phi}$. Thus diagram as in the left panel of Fig.~\ref{fig:clocksignal}, leading to a clock signal in the curvature spectrum, is absent here. One may wonder that a higher-order diagram, such as the one in Fig.~\ref{fig:model2cc}, could lead to a clock signal due to the resonance between $\sigma_{\rm bkg}$ and the intermediate $\tilde{\sigma}$ mode. However, it turns out not to be the case: the injection's frequency is the same as that of the intermediate $\delta \sigma$ and there is no resonant production, which is the same reason for the lack of clock signal in $cii$ in model 1 as discussed in the previous section.

\noindent \underline{\it Clock signal in the isocurvature spectrum.} Similar to model 1, there is still a clock signal in the isocurvature spectrum, due to the diagram in the right panel of Fig.~\ref{fig:clocksignal}. The computation is similar and the only change is $\tilde{\sigma}_{\rm bkg}$, which is Eq.~\eqref{eq:dim6featuresrho1} for model 2. We find that 
\begin{align}
 \left. \frac{\Delta P_i}{P_i} \right|_{\rm clock} &\simeq  \left. \frac{\Delta P_a}{P_a}\right|_{\rm clock}= \frac{-2 i}{\sqrt{1-x^2} f_I} \int_{-\infty}^0 \frac{d\tau}{(H\tau)^4} \bigg( \dot{u}_k^2 - \frac{k^2}{R^2} u_k^2 \bigg)\tilde{\sigma}_{\rm bkg} + \text{c.c.}~,\nonumber \\
&\simeq -\frac{x b V_{\phi 0}}{\dot{\tilde{\phi}}_0f_I  m_{\tilde{\sigma},\rm{eff}}} \sqrt{2\pi \mu_{\tilde{\sigma}}} \left(\frac{k}{k_r}\right)^{-3/2} \cos \left[ \mu_{\tilde{\sigma}} \ln \left(\frac{k}{k_r}\right) + \mu_{\tilde{\sigma}} +\frac{\pi}{4}  \right]  \label{eq:model2correctioniso2pt} 
\end{align}
where we take $\mu_{\tilde{\sigma}} \gg 1$ and $x \ll 1$ limits to simplify the final result. 
The amplitude (defined at $k=k_r$) is then: 
\begin{align}
 \left|\frac{\Delta P_i}{P_i}\right|_{\rm clock; amp} &= \frac{x b V_{\phi 0}}{\dot{\tilde{\phi}}_0 H f_I} \sqrt{\frac{2\pi}{\mu_{\tilde{\sigma}}}}  \nonumber \\
 &\simeq 1 \times \left( \frac{x}{0.1} \right)\left(\frac{\dot{\tilde{\phi}}_0}{(60 H)^2}\right) \left(\frac{40H}{f_I}\right)^{3/2} \left(\frac{b V_{\phi0}}{0.3\dot{\tilde{\phi}}_0^2} \right)\left(\frac{1}{\lambda}\right)^{1/4}~,
 \label{eq:model2Piclock}
\end{align}
where we use $\mu_{\tilde{\sigma}} \approx m_{\tilde{\sigma},\rm{eff}}/H \approx \sqrt{\lambda} f_I/ H$. This benchmark requires that the axion is only a subdominant fraction of dark matter with $\gamma/\theta_i \approx 10^{-3} \sqrt{\frac{\beta}{0.038}}$ to satisfy the current isocurvature bound in Eq.~\eqref{eq:isoconstraints}. In certain parameter space (e.g.~$x \gtrsim 0.1$ while fixing the other benchmark parameter values), the perturbative theory for the isocurvature spectrum could break down, and one needs to adopt a different method to compute its amplitude. 

Comparing the isocurvature clock signal amplitudes in Eq.~\eqref{eq:model1Piclock} for model 1 and Eq.~\eqref{eq:model2Piclock} for model 2, we note that they could be similar even though $x$ from the dimension-five operator (suppressed by one power of the cutoff scale) could be much larger than $q$ from the dimension-six operator (suppressed by two powers of the cutoff). This difference is compensated by different power dependence on other parameters, such as $\dot{\tilde{\phi}}_0/H^2 \simeq 60^2$. This serves as an example that the signal sizes are not entirely determined by the dimensions of the added operators.

\subsection{Three-point correlators}
\label{ssec:model2threepoint}

Analogous to model 1, the sharp feature also leads to a potentially large NG in $ccc$ via the diagram in Fig.~\ref{fig:cccclassical} and the result applies here as well. Similarly, the more interesting bispectrum in model 2 is $cii$, which we will discuss more in this section. The trilinear interactions of the perturbations read:
\begin{align}
        \mathcal{L}_{2}^{(3)} &\supset -\frac{1}{6} \frac{\partial^3 V_\phi}{\partial \tilde{\phi}^3}\delta\left(\tilde\phi -\frac{x}{\sqrt{1-x^2}}\tilde{\sigma}\right)^3 - \frac{1}{6}\frac{\partial^3 V_\sigma}{\partial \tilde{\sigma}^3}\delta\tilde{\sigma}^3 \nonumber\\   &+\left(\frac{1}{\sqrt{1-x^2}f_I}+\frac{\tilde{\sigma}_{\rm bkg}}{(1-x^2)f_I^2}\right)(\delta\tilde{\sigma}\delta\dot{a}^2-...)+\frac{x}{(1-x)^2 f_I}\bigg(\delta\dot{\tilde{\phi}}\delta\dot{\tilde{\sigma}}\delta\tilde{\sigma} -...\bigg)\nonumber\\
        & - \frac{x^2}{(1-x^2)^{3/2}f_I}\bigg( \delta\tilde{\sigma}\delta\dot{\tilde{\sigma}}^2 -...\bigg)~,
        \label{eq:trilinear}
\end{align}
where $\cdots$ represent the corresponding spatial derivative terms.

\begin{figure}[h!]
    \centering
    \includegraphics[width=10cm]{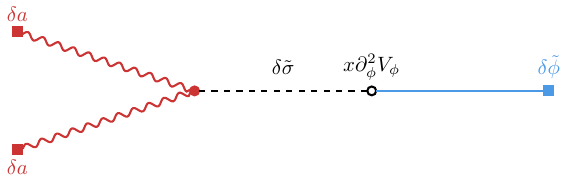}
    \caption{The leading diagram contributing to $cii$. The external point of inflaton is marked with blue squares while those of axions are marked with red ones. The solid blue, red wavy and dashed black lines represent the propagator for $\tilde{\phi}$, $a$, and $\tilde{\sigma}$ respectively. The empty dot denotes the interaction proportional to $x \partial^2_\phi V_{\phi}$ in the first line of Eq.~\eqref{eq:model2quadratic}. The filled red dot denotes the constant vertex between $\delta a$ and $\delta \tilde{\sigma}$.  }
    \label{fig:iicclassical}
\end{figure}

\noindent \underline{\it Mixed isocurvature-curvature bispectrum.} The leading diagram contributing to $cii$ is shown in Fig.~\ref{fig:iicclassical}. It contains the $\delta{\tilde{\sigma}}(\delta a)^2/f_I$ vertex in the second line of Eq.~\eqref{eq:trilinear} and one insertion of the $-x \partial^2_{\phi} V_\phi \delta \tilde{\phi} \delta \tilde{\sigma}$ vertex in the first line of Eq.~\eqref{eq:model2quadratic}.
This leading contribution from this diagram is
\begin{align}
    \langle\delta  a \delta a \delta\tilde{\phi}\rangle' &\supset \frac{-x u_{k_1} u_{k_2} u_{k_3}(\tau_{\rm end})}{(1-x^2)f_I}   \int^0_{-\infty} \frac{d\tau_1}{(H \tau_1)^4} \partial_\mu u^\ast_{k_1} \partial^\mu u^\ast_{k_2} v_{k_3}(\tau_1)
    \int^{\tau_1}_{-\infty} \frac{d\tau_2}{(H \tau_2)^4}\frac{\partial^2 V_\phi}{\partial \tilde{\phi}^2} u^\ast_{k_3} v^\ast_{k_3}(\tau_2)~.
\end{align}
The first integral at $\tau_1$ stands for the three-point vertex happening at late times, while the second integral is the mixing two-point vertex excited by the feature which happens at the earlier time $\tau_2$. The second integration is straightforward:
\begin{align}
 &\int^{\tau_1}_{-\infty} \frac{d\tau_2}{(H \tau_2)^4} \frac{\partial^2 V_\phi}{\partial \tilde{\phi}^2} u^\ast_{k_3} v^\ast_{k_3}(\tau_2)  \nonumber \\
&= \frac{\sqrt{\pi} b V_{\phi 0}(1+i)\left[-3+k_3 \tau_s (3i+2k_3 \tau_s)\right]}{8\dot{\tilde{\phi}}_0^2(-k_3 \tau_s)^{3/2}}e^{\pi \mu_{\tilde{\sigma}}/2} e^{i k_3 \tau_s} H^{(2)}_{i\mu_{\tilde{\sigma}}}(-k_3 \tau_s) \theta(\tau_1-\tau_s)~.
\end{align}
The three-point vertex at $\tau_1$ could be estimated in the same way as Eq.~\eqref{eq:threepointinf1}.

In the equilateral limit, where we apply the early-time expansion with $k/k_0\gtrsim \mu_\sigma$, the analytical approximation for $\langle\delta  a \delta a \delta \tilde{\phi}\rangle'$ to the leading order in $x$ reads
\begin{align}
   & \langle\delta  a \delta a \delta \tilde{\phi}\rangle' \simeq - \frac{3 i b V_{\phi 0} H^4 x e^{-3 i k/k_0}}{32 \dot{\tilde{\phi}}_0^2 k^6 f_I(1-x^2)}\bigg(\frac{k}{k_0}\bigg)~,
\end{align}
where we omit terms with lower powers of $k/k_0$. The corresponding dimensionless amplitude of the NG signal for $cii$ in the equilateral limit is then
\begin{align}
   \left| f_{\rm NL}^{cii}\right|\frac{A_i}{A_s} &\simeq \frac{3 \pi^3 x \beta H }{4 f_I \sqrt{A_s}} \times 
    \frac{b V_{\phi0}}{\dot{\tilde{\phi}}_0^2}~, \nonumber \\
   & \simeq 3.7 \left(\frac{40 H}{f_I}\right)\left(\frac{x}{0.1}\right) \left(\frac{\beta}{0.01}\right) \left(\frac{b V_{\phi0}}{0.3\dot{\tilde{\phi}}_0^2}\right).
\end{align}
Similar to the previous model in Section~\ref{sec:model1}, the observational prospect is still an open question as it depends on the oscillatory templates and multiple parameters.

\section{Model 3: $\partial_\mu \phi J^\mu + \chi \psi_L^\dagger \psi_R$ }
\label{sec:model3}
The first two models rely on a primordial inflationary feature to enhance the signals either in the two-point or the three-point correlators. In model 3, we will explore a different enhancement mechanism based on a chemical-potential type coupling between the inflaton and the PQ field, as well as the PQ field coupling to some other heavy spectator fermion fields. We find that in this model, there could also be novel correlated signals in several bispectra involving different numbers of the isocurvature modes. We also consider the cross power spectrum between the curvature and axion mode. 

In model 3, we add a different dimension-5 operator, $\partial_\mu \phi J_{\rm PQ}^\mu~,$ with $J_{\rm PQ}^\mu \equiv i(\chi^\dagger \partial^\mu \chi - \chi \partial^\mu \chi^\dagger)$, which again simultaneously respects the $U(1)_{\rm PQ}$ symmetry and the inflaton's shift symmetry, to the Lagrangian:
\begin{align}
{\cal L}_{\rm chem}&= -\frac{(\partial_\mu \phi)^2}{2} - |\partial_\mu \chi|^2 -   V(\phi) - \frac{\lambda}{2} \bigg(|\chi|^2 - \frac{f_a^2}{2} \bigg)^2 - i\frac{\kappa \partial_\mu \phi}{\Lambda} (\chi^\dagger \partial^\mu \chi - \chi \partial^\mu \chi^\dagger) \nonumber \\
& = - \frac{(\partial_\mu \phi)^2}{2} - \left[\left(\partial_\mu +i \frac{\kappa\partial_\mu \phi}{\Lambda}\right) \chi^\dagger\right]\left[\left(\partial^\mu -i \frac{\kappa\partial^\mu \phi}{\Lambda}\right) \chi \right] -   V(\phi) - \frac{\lambda}{2} \bigg(|\chi|^2 - \frac{f_a^2}{2} \bigg)^2 + \frac{\kappa^2 (\partial_\mu \phi)^2}{\Lambda^2}|\chi|^2 \nonumber \\
& = - \frac{(\partial_\mu \phi)^2}{2} - |\partial_\mu \tilde{\chi}|^2 -   V(\phi) - \frac{\lambda}{2} \bigg(|\tilde{\chi}|^2 - \frac{f_a^2}{2} \bigg)^2 +\frac{\kappa^2 (\partial_\mu \phi)^2} {\Lambda^2}|\tilde{\chi}|^2~,
\label{eq:self-coupling3}
\end{align}
where in the third line, we use the field redefinition as
\begin{equation}
    \tilde{\chi}\equiv e^{-i\frac{\kappa \phi}{\Lambda}} \chi~,
\end{equation}
to remove the kinetic mixing between $a$ and the inflaton.
After the spontaneous breaking of $U(1)_{\rm PQ}$, the radial and phase modes of $\tilde{\chi}= (f_I + \tilde{\sigma})/\sqrt{2} ~ {\rm exp} (i \tilde{a}/f_I)$ are related to the basis we start with, $\chi= (f_I + \sigma)/\sqrt{2} ~ {\rm exp} (i a/f_I)$, as
\begin{equation}
\tilde{\sigma} = \sigma~, ~~ \tilde{a} = a - z \phi~,
\label{eq:redef}
\end{equation}
with the dimensionless parameter $z$ defined as
\begin{equation}
   z\equiv \frac{\kappa f_I}{\Lambda}~.
\end{equation}
Both fields $\tilde{\sigma}$ and $\tilde{a}$ have canonically normalized kinetic terms. We identify $\tilde{a}$ as the massless axion mode, with the same Bunch-Davies initial condition as the inflaton. 
In this new basis, one could see that with the chemical-potential term alone, the axion $\tilde{a}$ is decoupled from the inflaton. Thus there is no mixed correlation, i.e., curvature-isocurvature spectrum and mixed bi-spectra. This is consistent with the discussions in the literature that $\partial_\mu \phi J^\mu$ coupling alone could not lead to any chemical potential for a scalar field (the axion in our case)~\cite{Wang:2019gbi,Bodas:2020yho,Sou:2021juh}.

We extend the model in Eq.~\eqref{eq:self-coupling3} by including a Dirac fermion $\psi$ that couples to the PQ scalar $\chi$:
\begin{equation}
    \mathcal{L}_3 ={\cal L}_{\rm chem} -  i \bar{\psi}\slashed{D} \psi  + \left(y\chi \psi_L^\dagger \psi_R + c.c.\right)~,
\end{equation}
where $D$ is the covariant derivative, and $\psi_L, \psi_R$ are the two chiral Weyl components of $\psi$: $\psi = \begin{pmatrix}
\psi_L  \\
\psi_R 
\end{pmatrix}$.
The above Lagrangian will be invariant under the global PQ symmetry if the component Weyl fermions are also charged under $U(1)_{\rm PQ}$ with opposite charges. Under a PQ rotation by an angle $\theta$, the charged matter transforms as 
\begin{equation}
    \chi \to \chi e^{i \theta}~,~\psi_L \to \psi_L e^{i \theta/2}~,~\psi_R \to \psi_R e^{-i \theta/2}~. 
\end{equation}
After the PQ symmetry breaking, $\psi$ acquires an effective mass 
\begin{equation}
m_\psi = \frac{y f_I}{\sqrt{2}} ~.
\end{equation}
Performing a chiral rotation of the fermions to remove the phase in the Yukawa term, the fermion kinetic term yields:
\begin{equation}
     i \bar{\psi}\slashed{D} \psi \xrightarrow[\psi_R \to e^{i a/(2f_I)}\psi_R]{\psi_L \to e^{-i a/(2f_I)}\psi_L} \frac{\partial_\mu a}{2 f_I} \bar{\psi}  \gamma^\mu \gamma^5\psi + ...~,
\end{equation}
where we omit irrelevant terms for the CC signals. The heavy fermions $\psi$ above could be charged under the SM color group to induce a coupling between the axion and the SM gluon as needed in the KSVZ axion model~\cite{Kim:1979if,Shifman:1979if}. In the presence of $\partial_\mu \phi J^\mu$, we identify $\tilde{a}$ as the axion with its homogeneous background satisfying $\dot{\tilde{a}}_0=0$ during inflation. This implies $a$ rolls during inflation with a speed $\dot{a}_0 = z \dot{\phi}_0$. Then its coupling to the non-conserved axial current $J^\mu_5=\bar{\psi}  \gamma^\mu \gamma^5 \psi$ gives rise to a chemical potential for the fermion
\begin{equation}
\mu_c \equiv  \frac{z \dot{\phi}_0}{2 f_I}~. 
\end{equation}
When the chemical potential $\mu_c \gtrsim m_\psi \simeq y f_I/\sqrt{2}$, $\psi$ with a particular chirality would be produced efficiently from the inflationary background, leading to sizable NG signals~\cite{Chen:2018xck, Sou:2021juh}. Using the redefined basis in Eq.~\eqref{eq:redef}, the couplings between the fermion and the scalars read 
\begin{equation}
    \frac{\partial_\mu(z\phi+\tilde{a})}{2 f_I} \bar{\psi}\gamma^\mu \gamma^5 \psi~.\label{eq:fermioncoupling}
\end{equation}

One naturalness requirement is that the Yukawa coupling $y$ should satisfy
\begin{equation}
    \frac{y^4}{16\pi^2} \lesssim \lambda~,
\end{equation}
in order not to generate a large radiative correction to the quartic coupling of the PQ field. Another constraint is on the radiative correction to the PQ field mass squared term from the fermion loop,
\begin{equation}
 \frac{y^2 \Lambda^2}{16\pi^2} \lesssim \lambda f_a^2~. 
\end{equation}
This could be achieved if the previous condition on $y$ is satisfied and $\Lambda \lesssim 4 \pi f_a$.
In addition, $z \ll 1$ so that $\kappa^2 (\partial_\mu \phi)^2 |\tilde{\chi}|^2/\Lambda^2$ will not lead to a large negative contribution to the inflaton kinetic term. Finally, to avoid the last term in Eq.~\eqref{eq:self-coupling3} restoring the PQ symmetry during inflation, we also need
\begin{equation}
    \kappa^2 \frac{\dot{\phi}_0^2}{\Lambda^2} \lesssim \frac{\lambda}{2} f_a^2  \Rightarrow f_a^2 \gtrsim \sqrt{\frac{2}{\lambda}} z\dot{\phi}_0~.
\end{equation}
In practice, $f_I \lesssim f_a$ is possible with a certain level of fine-tuning. In general we expect $f_I$ to be of $\mathcal{O}(\sqrt{\dot{\phi}_0})$ or higher. 

\begin{figure}[h!]
    \centering
    \includegraphics[width=\textwidth]{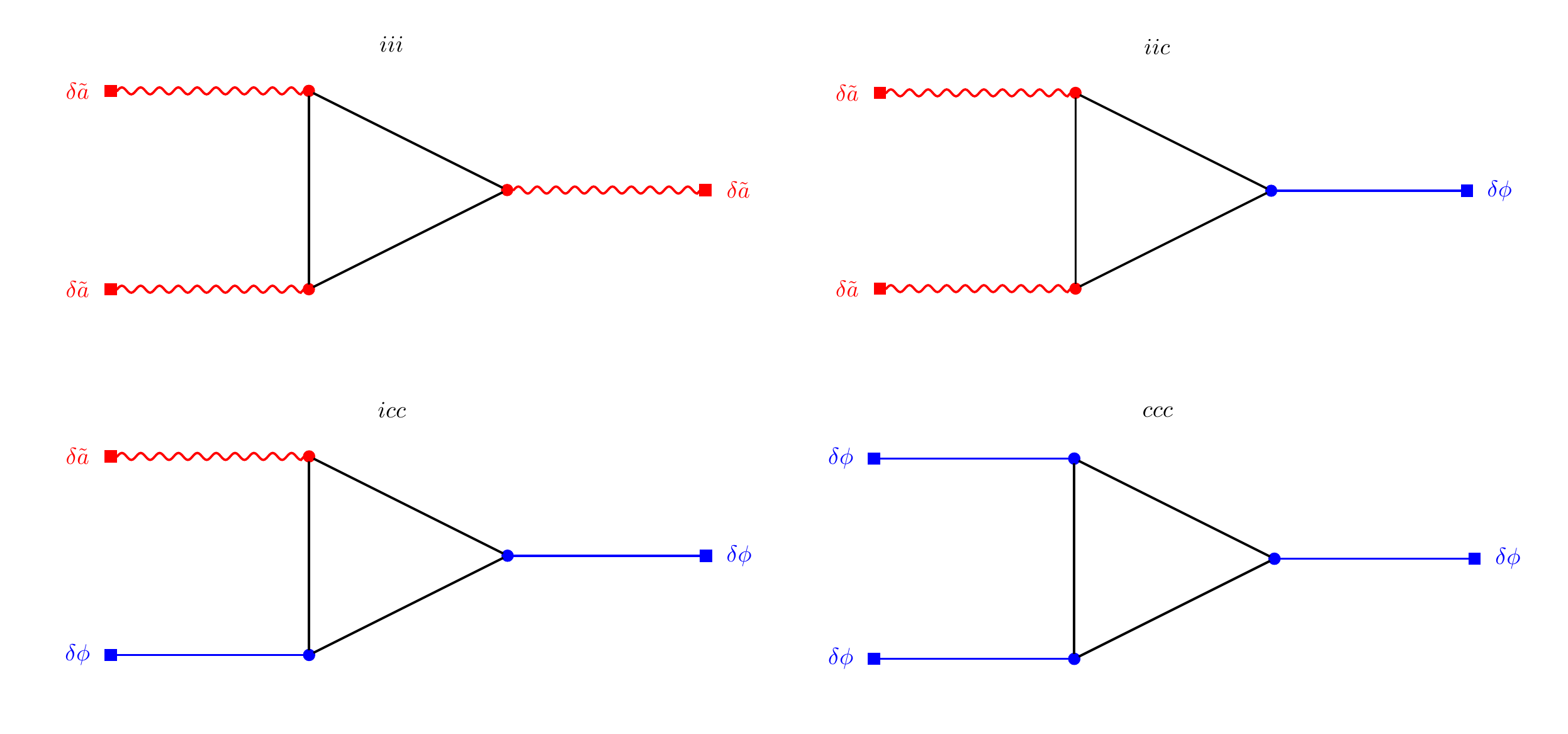}
    \caption{The fermion loop diagrams for $iii$, $iic$, $icc$ and $ccc$ correlators. The red (blue) squares indicate the external axion (inflaton) mode. The red (blue) dots represent the coupling of fermion to axion (inflaton) in Eq.~\eqref{eq:fermioncoupling}. The red wavy, blue, and black lines represent the axion, inflaton, and fermion propagators, respectively. }
    \label{fig:fermionloop}
\end{figure}

Since the fermion couples to both $\tilde a$ and $\phi$ as shown in Eq.~\eqref{eq:fermioncoupling}, its loop introduces all kinds of CC bispectra signal in terms of curvature and isocurvature, as shown in Fig.~\ref{fig:fermionloop}. Since the coupling of the inflaton and $\tilde{a}$ to the fermions is proportional, we expect the bispectra signals to be related to each other in a simple way. For the pure isocurvature bispectra ($iii$), its NG parameter $f_{\rm NL}^{iii}$ can be estimated using the late-time expansion of fermion modes~\cite{Chen:2018xck,Hook:2019zxa}:
\begin{align}
    \left|f_{\rm NL}^{iii}\right| \frac{A_i^2}{A_s^2} &\simeq \frac{N_c N_\psi \beta^{3/2}}{6\pi\sqrt{A_s}} \bigg( \frac{H}{2 f_I}\bigg)^3 \left(\frac{m_\psi}{H}\right)^3 
    \frac{\mu_c^2\sqrt{m_\psi^2+\mu_c^2}}{H^3}\nonumber\\
    &\times \frac{e^{\pi \mu_c/H} \Gamma \left(-i\sqrt{m_\psi^2+\mu_c^2}/H\right)^2\Gamma \left(2i\sqrt{m_\psi^2+\mu_c^2}/H\right)^3}{2\pi \Gamma \left[i\left(\sqrt{m_\psi^2+\mu_c^2}+\mu_c\right)/H\right]^3\Gamma\left[1+i\left(\sqrt{m_\psi^2+\mu_c^2}-\mu_c\right)/H\right]}~,
\end{align}
where we set $N_c$ and $N_\psi$ are $\psi$'s number of color and flavor, respectively. For the QCD axion that couples to vector-like quarks, $N_c=3$ while $N_\psi\geq 1$. 

\begin{figure}[h!]
    \centering
    \includegraphics[width=7.5 cm]{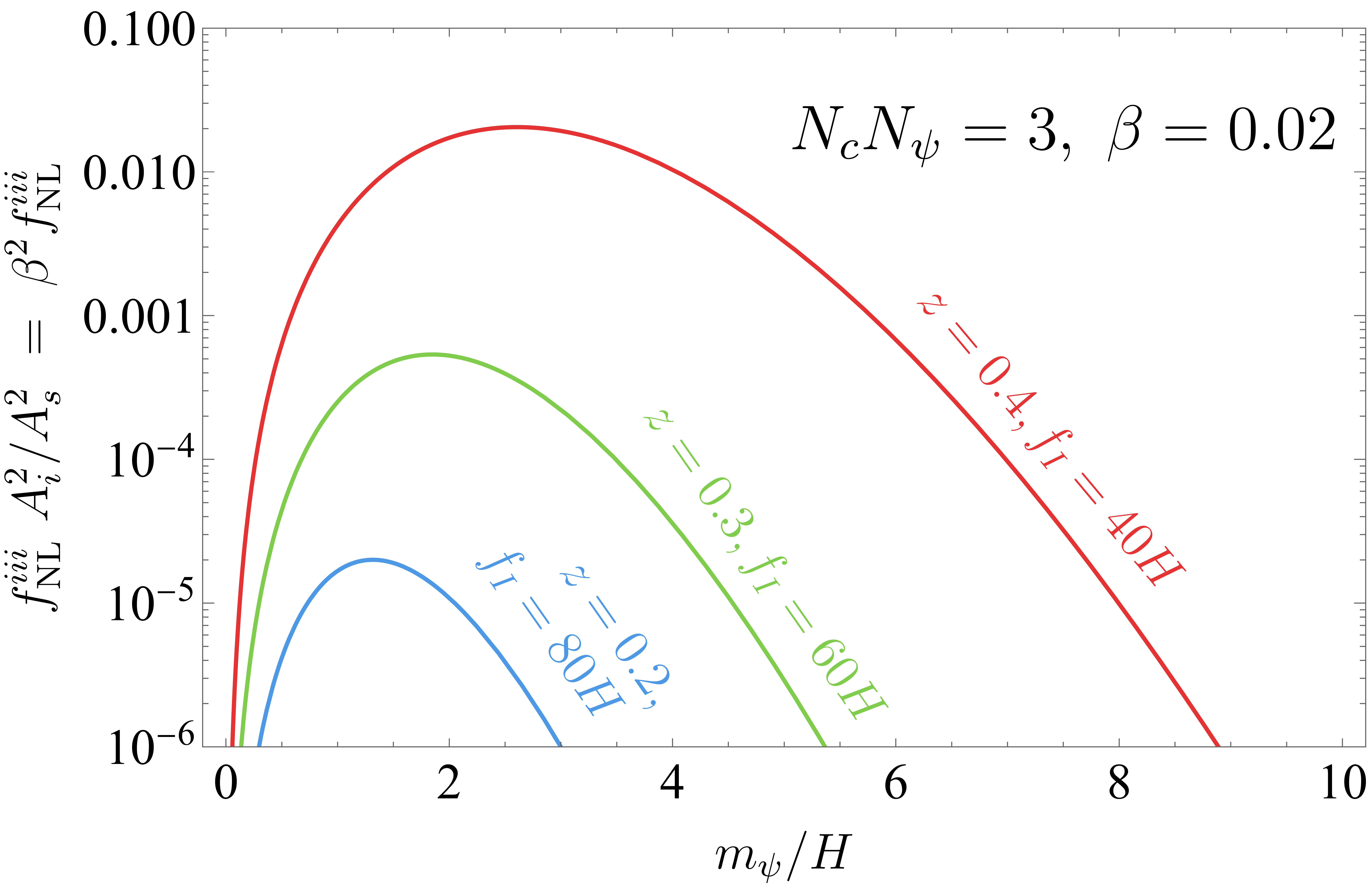}
    \includegraphics[width=7.5 cm]{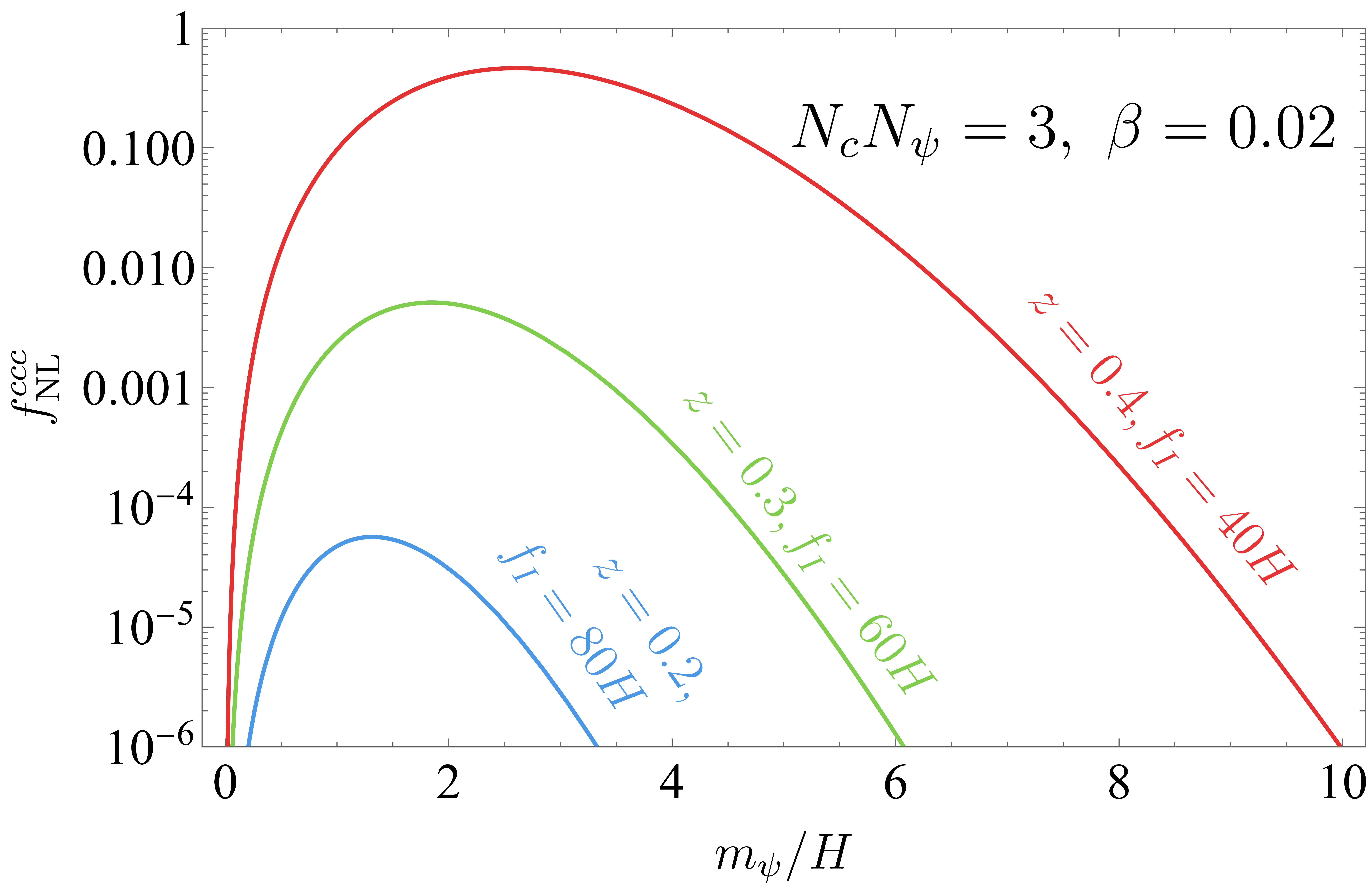}
    \caption{The $f_{\rm NL}$ of pure isocurvature (left) and curvature (right) CC signals induced by the fermion loop. In both plot, we have the benchmark total fermion number $N_cN_\psi =3$ and $\beta\simeq A_i/A_s =0.02$ to satisfy the constraint in Eq.~\eqref{eq:isoconstraints}. }
    \label{fig:fnlfermion}
\end{figure}

For the other (mixed) bispectra, we have correlated signals with the same mass parameter and related amplitudes:
\begin{align}
        f^{iic}_{\rm NL} A_i A_s&\simeq 2f^{cii}_{\rm NL} A_i A_s = z \beta^{-1/2} f_{\rm NL}^{iii} A_i^2~,\nonumber\\
        f^{cci}_{\rm NL}A_i A_s&\simeq 2f^{icc}_{\rm NL}A_i A_s = z^2 \beta^{-1} f_{\rm NL}^{iii} A_i^2~,  \nonumber \\
        f^{ccc}_{\rm NL} A_s^2 &= z^3 \beta^{-3/2}f^{iii}_{\rm NL} A_i^2~.
\end{align}
Here the superscripts of $f_{\rm NL}$ are ordered, with the first index corresponding to the mode with the lowest momentum in the squeezed limit. Depending on the sizes of $z$ and $\sqrt{\beta}$, the CC signals in each mixture vary. Fig.~\ref{fig:fnlfermion} shows quantitative results of $f_{\rm NL}^{iii}$ and $f_{\rm NL}^{ccc}$. One could see that the amplitude of the curvature bispectrum is generally small, but in some special parameter space, $f_{\rm NL}$ becomes much greater than ${\cal O}(0.01)$. Note that, compared to the first two models, the non-Gaussianities here are scale-invariant. Also the magnitudes of $f_{\rm NL}$ are generally smaller due to loop effects. Therefore it would be very difficult to observe such non-Gaussian signals, although in general the constraints and detection prospects of these types of non-Gaussianities involving isocurvature external lines remain as open questions.

On the other hand, the fermion loop also introduces a mixing between $\Tilde{a}$ and $\phi$, and thus the correlation between curvature and isocurvature modes at the one-loop level. However, since the scale-invariance is unbroken in the two-point functions, the fermion loop only appears as the local part, and the mixing is suppressed by $m_\psi^{-2}$. A quick estimation of the mixed spectrum from the fermion loop reads:
\begin{equation}
(2\pi)^2 A_{is} \simeq z\sqrt{\beta}\bigg(\frac{H}{\dot{\phi}_0} \bigg)^2   \frac{H^4}{16\pi^2 m_\psi^2} \frac{1}{(2f_I)^2} e^{-2\pi\left(\mu_c-\sqrt{\mu_c^2+m_\psi^2}\right)}~, 
\end{equation}
and the corresponding mixing angle is
\begin{equation}
    \cos\Delta \simeq \frac{A_{is}}{\sqrt{A_s A_i}} \simeq\frac{A_{is}}{A_s\sqrt{\beta}} \lesssim 10^{-5},
\end{equation}
when $m_\psi \gtrsim H$. Such a small correlation is thus very challenging for observations and is unlikely to cause more stringent bounds on $\beta$. 

One could consider a similar model with the axion coupling to heavy spectator gauge bosons instead of fermions. But that type of model requires more ingredients to generate a sufficiently large CC signal. We leave the explanation and estimates of the signals to Appendix~\ref{app:gaugeboson}.

\section{Conclusions and outlook}
\label{sec:outlook}

In this article, we point out that if a light axion is present during inflation and becomes even only a small fraction of dark matter later, its associated isocurvature fluctuations could result in a profusion of novel cosmological correlators, depending on the interaction between the PQ field and the inflaton. These correlators are either two-point or three-point functions involving different combinations of curvature and isocurvature modes. The corresponding signals include correlated classical clock signals, taking the form of $\sin(\log k)$ with $k$ the wavenumber, in both the curvature and isocurvature power spectra; and the cosmological-collider type NG in mixed curvature-isocurvature or pure isocurvature three-point functions, which could be correlated as well. With the help of a classical inflationary feature or chemical-potential enhancement, these signals could be sizable and observable in the future, even in the presence of current strong constraints on the overall amplitude of the isocurvature power. They encode a wealth of information about the interplay between the PQ breaking dynamics and inflation at a very high energy scale, which could be challenging to be probed otherwise. In some models, the signals could be combined to infer the range of the inflationary Hubble scale, an important parameter which suffers from a lack of observables, independent of the tensor mode. 

Our paper serves as an early step to explore the application of the booming cosmological collider and primordial standard clock physics to explore the axion dynamics during the inflationary epoch in the very early universe. There exists more to study along the direction:
\begin{itemize}
\item The paper is based on three simple EFTs, each containing only one high-dimensional operator coupling the PQ field to the inflaton. A general EFT could contain several operators simultaneously. New diagrams could be present and lead to different patterns of the observables. 
\item In our study, we do not intend to solve the QCD axion isocurvature problem and ameliorate the tension between the high-scale inflation scenarios and QCD axion dark matter. There exist quite a few models to address this problem by including specific types of interactions between the inflaton and the PQ field~\cite{Linde:1991km, Jeong:2013xta, Choi:2014uaa, Chun:2014xva, Fairbairn:2014zta, Nakayama:2015pba, Harigaya:2015hha, Kearney:2016vqw}. It would be interesting to see whether these models could still give rise to cosmological collider or primordial classical clock signals in the parameter space where the QCD axion isocurvature problem is solved. 
\item On the observational side, the signals we point out in the paper require new templates in the data analyses. The most outstanding example is that in our first two models, the leading component in the isocurvature power spectrum could be an oscillatory clock, which may not be captured by the current Planck template.   
\item It would be informative to study the constraints and forecast the future experimental prospects of the oscillatory isocurvature power spectrum and mixed curvature-isocurvature power spectrum, as well as those of the scale- or shape-dependent oscillatory mixed curvature-isocurvature bispectra. 
\end{itemize}

\section*{Acknowledgement}
We thank Arushi Bodas, Matteo Braglia, Anson Hook, Soubhik Kumar, Lucas Pinol, Matt Reece, Raman Sundrum, Xi Tong, Benjamin Wallisch, Zekai Wang, Yi-Peng Wu, and Zhong-Zhi Xianyu for useful comments and feedback. JF and LL thank the physics department of Harvard University for hospitality when this work is conducted. JF and LL are supported by the NASA grant 80NSSC22K081 and the DOE grant DE-SC-0010010.

\appendix

\section{Mixed bi-spectra from massive gauge boson loops}
\label{app:gaugeboson}
In this appendix, we will discuss another possible scenario with CC signals boosted by chemical-potential coupling. 
In addition to fermions, one could also have the axion interact with the gauge bosons. Consider the following interactions on top of $\mathcal{L}_{\rm chem}$ in Eq.~\eqref{eq:self-coupling3}, similar to~\cite{Wang:2020ioa}:
\begin{equation}
- |\text{D}_\mu \Sigma|^2 -V(\Sigma) -\frac{1}{4} F^{\mu \nu} F_{\mu \nu} - \frac{|\partial_\mu \phi|^2}{\Lambda_{\Sigma}^2}|\Sigma|^2 - \frac{a}{4 \Lambda_f} F^{\mu \nu} \tilde{F}_{\mu \nu}~.
\label{eq:self-coupling4}
\end{equation}
Here we introduce a dark Higgised Abelian gauge group, $U(1)_d$. The complex scalar charged under $U(1)_d$, $\Sigma$, is the Higgs field that acquires a VEV and makes the $U(1)_d$ gauge boson $A$ massive. $F$ is the field strength of $U(1)_d$. We consider the Higgs field coupling to the inflaton via the dimension-six operator $(\partial_\mu \phi)^2|\Sigma|^2$. This mixing between the inflaton and the Higgs field will amplify the NG signals as shown in~\cite{Wang:2020ioa}. The cutoff scales, $\Lambda_f$ and $\Lambda_\Sigma$, determine the interaction strength, which are not necessarily equal in general. $\frac{a}{4 \Lambda_f} F^{\mu \nu} \tilde{F}_{\mu \nu}$ could be generated via integrating out a heavy fermion loop with the fermions coupled to the PQ field $\chi$, as in KSVZ-type axion model~\cite{Kim:1979if, Shifman:1979if}. Then $\Lambda_f$ is related to the PQ breaking scale $f_I$ as $\Lambda_f \simeq 2\pi f_I/(a_d \alpha_d)$, where $a_d$ is the coefficient of the mixed $U(1)_{\rm PQ}$ and $U(1)_d$ anomaly, and $\alpha_d$ the $U(1)_d$ fine-structure constant. 

Schematically the potential of $\Sigma$ can be written as:
\begin{equation}
    V(\Sigma) = -m_\Sigma^2 |\Sigma|^2 +\lambda |\Sigma|^4~,
\end{equation}
with $m_\Sigma^2, \lambda>0$.
During inflation, the coupling between the inflaton and Higgs field in Eq.~\eqref{eq:self-coupling4} introduces an additional contribution to the effective mass squared of $\Sigma$, $\dot{\phi}_0^2/\Lambda_\Sigma^2$. Using the parameterization $\Sigma\equiv (\sigma/\sqrt{2})e^{i \pi}$, the VEV of the Higgs field is given by $\langle \sigma\rangle^2 = (\dot{\phi}_0^2/\Lambda_\Sigma^2+ m_\Sigma^2)/\lambda = m_\sigma^2/(2\lambda)$. The gauge boson mass $m_A$ during inflation is given by $m_A=g\langle \sigma\rangle$, with $g$ the $U(1)_d$ gauge coupling.

After the field redefinition in Eq.~\eqref{eq:redef}, the coupling between the heavy gauge field and scalars reads 
\begin{equation}
    \frac{(z\phi+\tilde{a})}{\Lambda_f} F\tilde{F}~,~~z\equiv\frac{\kappa f_I}{\Lambda}~.
\end{equation}
The classical rolling of the inflaton will lead to a chemical potential $\mu$ for the heavy gauge boson, 
\begin{equation}
    \mu \equiv z\dot{\phi}_0/\Lambda_f~.
\end{equation}
We consider both $m_A$ and $\mu$ to be $\gtrsim H$. Since a large $m_A$ leads to an exponential suppression, which has to be compensated by the chemical potential for an observable CC signal, we need 
\begin{equation}
 m_A f_I \lesssim \mu f_I = \frac{z \dot{\phi}_0}{\Lambda_f}\times f_I \simeq \frac{z (a_d \alpha_d) }{4\pi^2 \sqrt{A_s}}H^2 \simeq 5.4 \times{10^2} z (a_d \alpha_d)  H^2~,
\end{equation}
where we use $A_s= H^4/(2 \pi \dot{\phi}_0)^2$, and $\Lambda_f \simeq 2 \pi f_I/(a_d \alpha_d)$. 
Therefore, $f_I$ and $m_A$ cannot take very large values simultaneously even when $z, a_d \alpha_d \simeq 1$.

\begin{figure}
    \centering
    \includegraphics[width=10cm]{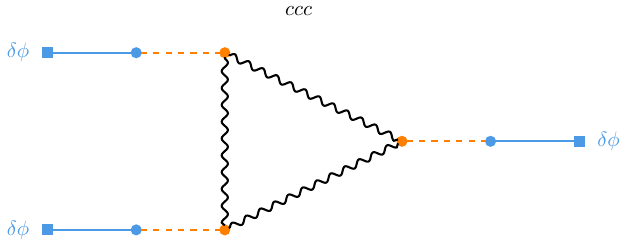}
    \caption{Diagram contributing to $ccc$ through a gauge boson loop. The solid blue, dashed orange, and wavy black lines are the propagators of the inflaton, the Higgs $\sigma$ that mixes with the inflaton, and the gauge boson, respectively. The filled dots indicate constant couplings and the squares indicate the external points. }
    \label{fig:gaugeloop1}
\end{figure}

Following~\cite{Wang:2020ioa}, we will estimate the amplitudes for different types of bispectra, using simple power counting rules.
The $ccc$ bispectrum has already been estimated in~\cite{Wang:2020ioa}. The key observation is that the one-loop diagram with no insertion of the inflaton and Higgs mixing is highly suppressed. On the other hand, for diagrams with $\sigma$ mixing with the inflaton, e.g., Fig.~\ref{fig:gaugeloop1}, the CC signal is amplified since the mixing is of order $\dot{\phi}_0 \langle \sigma \rangle$ and both factors could be large compared to $H$. The oscillating part of $ccc$ reads, when $\tilde{\mu} =\mu/H, \tilde{\nu} \equiv \sqrt{(m_A/H)^2 -1/4}\approx m_A/H > 1$: 
\begin{align}
\langle \zeta \zeta \zeta\rangle' = (2\pi)^4 f^{ccc}_{\rm NL} A_s^2 & \sim \frac{H^3}{\dot{\phi}_0^3} \frac{1}{16 \pi^2}\bigg( \frac{\dot{\phi}_0 \langle \sigma \rangle}{\Lambda_\Sigma^2} \frac{1}{m_\sigma^2} 2g^2 \langle\sigma\rangle \bigg)^3 \frac{H^2}{m_A^2} \bigg(\frac{m_A}{H}\bigg)^{3/2} \times e^{p}~,\nonumber \\
&= 4 \pi^4 A_s^3 \bigg(\frac{\dot{\phi}_0^2 }{\langle \sigma \rangle^2 \Lambda_\Sigma^2} \bigg)^3 \bigg(\frac{m_A}{H}\bigg)^{11/2}\times e^p~,\nonumber \\
& =4 \pi^4 A_s^3 u^{-3} \bigg(\frac{m_A}{H}\bigg)^{11/2} e^p~, \label{eq:leadingbispectrum} \\
p &= 6\pi(\tilde{\mu}-\tilde{\nu})~, ~~ {\rm when} ~ \tilde{\mu} \geq \tilde{\nu}~,  \nonumber \\
 &= 2 \pi (\tilde{\mu}- \tilde{\nu})~, ~~ {\rm when} ~ \tilde{\mu} < \tilde{\nu}~.  \nonumber
\end{align}
In the equations above, we have $u\equiv 1+\frac{m_\Sigma^2 \Lambda_\Sigma^2}{\dot{\phi}_0^2}= \frac{2\lambda\langle \sigma \rangle^2 \Lambda_\Sigma^2}{\dot{\phi}_0^2 } \sim 1$ when the bare mass $m_\Sigma$ is small. In the first line, all the factors are from simple power counting rules except for an extra power $(m_A/H)^{3/2}$, which comes from an explicit full loop integration with late-time expansion and is not covered by the power counting rules~\cite{Wang:2020ioa}. We require $\tilde{\mu} \sim \tilde{\nu}$ and the exponential factor could not be very large. Otherwise, the computation may not hold as it enters the non-perturbative regime. 

On the other hand, replacing the external leg from $\phi$ to $\tilde{a}$ will produce bispectra signals involving isocurvature modes, such as the $icc$ type. In this case, both the constraint on $\beta$ in Eq.~\eqref{eq:isoconstraints}, and the small $aF\tilde{F}$ coupling will suppress the signal. In particular, one needs to replace vertexes in Eq.~\eqref{eq:leadingbispectrum} for each $\tilde{a}$ external leg attached: 
\begin{equation}
\bigg(\frac{\dot{\phi}_0   \langle \sigma \rangle}{\Lambda_\Sigma^2} \frac{1}{m_\sigma^2} 2g^2 \langle\sigma\rangle \bigg)\simeq \frac{m_A^2}{u\dot{\phi}_0} \to \frac{H}{\Lambda_f}~.
\end{equation}
In addition, when converting $\delta \tilde{a}$ to the DM isocurvature perturbation $S_{d}$, one needs to replace the curvature normalization by
\begin{equation}
\frac{H}{\dot{\phi}_0} \to \frac{2 \gamma}{f_I\theta_i} \simeq \sqrt{\beta}\frac{H}{\dot{\phi}_0}~.
\end{equation}
Note that since one Higgs-gauge vertex $\sigma A^2$ is replaced by $aF\tilde{F}$, the extra power dependence $(m_A/H)^{3/2}$ from the loop integration may change as well. We will not explore it further here. Assuming that the power dependence doesn't change, we estimate
\begin{align}
    \frac{f^{icc}_{\rm NL} A_s A_i}{f^{ccc}_{\rm NL} A_s^2} &\simeq  \sqrt{\beta}\frac{H}{\Lambda_f} 
 \frac{u \dot{\phi}_0}{m_A^2}  \simeq u \sigma \sqrt{\beta}\frac{H}{m_A z}~, \\
 \sigma & \equiv \frac{\mu}{m_A} \sim 1~. 
\end{align}
Since $\beta \lesssim 0.04$ and $m_A/H \lesssim 10$, we expect the $icc$ type non-Gaussianity to be suppressed by ${\cal O} (10^{-2})$, compared to $ccc$. 

The same gauge boson loop will also introduce two-point mixing between adiabatic and isocurvature modes. Equivalently, the correlation between $A_s$ ad $A_i$, parameterized by a mixing angle $\cos\Delta$, will be nonzero. However, since the scale-invariance is unbroken in two-point functions, the gauge boson loop only appears as the local part, and the mixing is suppressed by $m_A^{-2}$. A quick estimation of the mixed spectrum from the loop reads:
\begin{equation}
(2\pi)^2 A_{is} \simeq \sqrt{\beta}\bigg(\frac{H}{\dot{\phi}_0} \bigg)^2   \frac{H^4}{16\pi^2 m_A^4} \frac{m_A^2}{u\dot{\phi}_0} \frac{H}{\Lambda_f} e^{\tilde{p}} \simeq \pi^2 \frac{A_s^2 H \sigma \sqrt{\beta}}{m_A u z} 
\times e^{\tilde{p}}~, 
\end{equation}
with $\tilde{p}= 4 \pi (\tilde{\mu}-\tilde{\nu})$ when $\tilde{\mu} \geq \tilde{\nu}$ and 0 otherwise. 
and correspondingly
\begin{equation}
    \cos\Delta \simeq \frac{A_{is}}{\sqrt{A_s A_i}} \simeq\frac{A_{is}}{A_s\sqrt{\beta}} \simeq \frac{ A_s H}{4 m_A z}\times e^{\tilde{p}}~,
\end{equation}
where in the last step, we take $\sigma, u \sim 1$ and $\cos \Delta$ is highly suppressed.

The Higgs-mixing mechanism may also be used to enhance the bi-spectrum involving the isocurvature modes. Consider replacing the relevant term in Eq.~\eqref{eq:self-coupling4} as:
\begin{equation}
 \frac{|\partial_\mu \phi|^2}{\Lambda_\Sigma^2}|\Sigma|^2 \to    \frac{|\partial_\mu \chi|^2}{\Lambda_\Sigma^2}|\Sigma|^2 ~.
\end{equation}
After field redefinition of $\tilde{a}$, the term lead to the effective coupling $\frac{(\partial (\tilde{a}+z \phi))^2}{4\Lambda_\Sigma^2} \sigma^2$ and thus the mixing between $\sigma$ and $\dot{\tilde{a}}$ is about $\frac{z\langle \sigma\rangle\dot{\phi}_0}{2\Lambda_\Sigma^2} \sigma\dot{\tilde{a}}$. One can thus estimate the enhanced $iii$ type of bispectrum signal as:
\begin{align}
(2\pi)^4f^{iii}_{NL} A_s^2 & \sim \bigg(\frac{2 r}{f_I \theta_i}\bigg)^3 \frac{1}{16 \pi^2}\bigg( \frac{z\dot{\phi}_0 \langle \sigma \rangle}{\Lambda_\Sigma^2} \frac{1}{m_\sigma^2} 2g^2 \langle\sigma\rangle \bigg)^3 \frac{H^2}{m_A^2} \bigg(\frac{m_A}{H}\bigg)^{3/2} e^p~ \nonumber \\
& \simeq 4\pi^4 A_s^3 \beta^{3/2} \left(\frac{z}{u}\right)^3 \bigg(\frac{m_A}{H}\bigg)^{11/2} e^p~, 
\end{align}
which could still be observable despite the suppression factor of $\beta^{3/2} z^3$. 

One could also consider a scenario in which both Higgs-inflaton mixing and Higgs-axion mixing are present, in which the mixed-spectra won't be suppressed by additional powers of $H/m_A$, compared to the $ccc$ type.

\bibliography{Ref}

\end{document}